\documentclass[12pt]{article}
\usepackage{lmodern}
\usepackage[top=1in, bottom=1in, left=1in, right=1in]{geometry}

\usepackage{xcolor}

\usepackage{pdflscape}
\usepackage{graphicx}
\usepackage{booktabs}
\usepackage{tabularx}
\usepackage[labelfont=bf]{caption}
\usepackage{multirow}
\usepackage{array}
\usepackage{ragged2e}
\usepackage{setspace}

\usepackage{amsmath}
\usepackage{bm}
\usepackage{amssymb, amsfonts}
\usepackage{amsthm}
\usepackage{mathrsfs}

\usepackage{authblk}
\usepackage{natbib}
\usepackage[title]{appendix}
\usepackage[colorlinks=true,linkcolor=blue,urlcolor=blue,citecolor=blue]{hyperref}

\def\sym#1{\ifmmode^{#1}\else\textsuperscript{#1}\fi}

\title{\textbf{Using Language Models to Predict Group Experiment Results}\thanks{The author marked with an asterisk (*) is the corresponding author. Please direct comments or questions to Xun Pang at xpang@pku.edu.cn. This manuscript employed large language models for grammar correction and proofreading.}}
\author[1]{Yifei Zhu}
\author[2]{Jingtian Hu}
\author[4]{Yinzhi Lin}
\author[3]{Zhenhua Tu}
\author[3,*]{Xun Pang}
\affil[1]{Department of Politics and Public Administration, The University of Hong Kong}
\affil[2]{Department of International Relations, School of Social Science, Tsinghua University}
\affil[3]{School of International Studies, Peking University}
\affil[4]{School of Government, Peking University}
\date{\today}

\begin{document}
\maketitle

\begin{abstract}
Can large language models (LLMs) predict how human groups respond to experimental treatments? We compare how people and simulated participants respond to the same treatments in \citet{Kertzer_2022_Hawkish}'s foreign-policy experiments. Silicon subjects recover human treatment directions across three experiments spanning individual and group settings, with correspondence varying across tasks and advisory structures. Calibration suggests that silicon samples, like biased human samples, can be informative even when they misestimate the size of a treatment effect. These findings support using LLM simulations to identify treatments worth testing in human group experiments.

\end{abstract}

\newpage
\doublespacing

\section{Introduction}
Group experiments can reveal how advisers share information and how decision rules shape a collective choice \citep{StasserTitus_1985_Unshared,JostKertzerMinSchub_2024_Advisers,Kertzer_2022_Hawkish}. Yet testing each new scenario takes time and money because researchers must recruit and coordinate enough groups for a comparison. Large language models (LLMs) offer a lower-cost way to screen designs by simulating deliberation among model-generated participants, or \textit{silicon subjects} \citep{Park_2024_Generative,cui2025largescale}. We ask whether these simulations reproduce both how human groups respond to a treatment and how much their members disagree after discussion.

We ask whether silicon subjects can predict the direction of human treatment effects well enough to help researchers choose which treatments to test. Drawing on research evaluating convenience samples in survey experiments \citep{berinsky2012mturk,mullinix2015generalizability,Coppock_2019_GeneralizingMTurk}, we assess \textit{behavioral correspondence} by comparing how the same treatment changes decisions in human and simulated experiments. A simulation might correctly predict that framing a choice as a loss increases risk-taking on average yet miss which respondents are most sensitive to the frame.

We study this approach using survey experiments in international relations (IR), which identify the microfoundations of foreign-policy decision-making \citep{Hafner_2017,KertzerTingley_2018_PoliticalPsychology,KertzerRenshon_2022_Elites}. Four prominent experiments fielded across purposively varied national contexts produced stable treatment directions, a pattern associated with limited treatment-effect heterogeneity \citep{BassanNygateRenshonWeeksWeiss_2024_Generalizability}. These findings make prominent microfoundational IR experiments a useful test bed for language-model-based discovery. They offer established human benchmarks for exploring candidate treatment patterns, scope conditions, and group configurations.

We evaluate behavioral correspondence using \citet{Kertzer_2022_Hawkish}'s hawkish-bias experiments as the human benchmark. Across three tasks involving prospect theory, intentionality bias, and reactive devaluation, participants decide individually, deliberate as peers, or advise a designated leader. We reproduce these roles in the simulations. In peer groups we use the median final response as the group decision. In hierarchical groups the leader makes the final choice after hearing from advisers. Within each setting we compare the human and silicon treatment effects. For groups matched on membership and treatment, we also compare how widely members' final reports differ. Respondent covariates allow separate checks on treatment heterogeneity.

Silicon subjects recover the human treatment-effect classifications in all three individual experiments. In groups, human and silicon treatment effects point in the same direction in five of six comparisons. The calibration results suggest that silicon subjects, like respondents in a convenience sample, can provide useful information about treatment effects. Adding their predictions sometimes improves the estimate from a human sample that overrepresents certain groups.

The article makes three contributions. First, it defines behavioral correspondence as an estimand-specific framework for evaluating silicon subjects and separates contextual correspondence from sample correspondence. Second, it extends this evaluation to deliberating groups. Comparing both collective choices and disagreement among members gives researchers two criteria for judging whether a simulation can inform the design of a human group experiment. Third, it uses calibration to test whether biased silicon responses still carry information about human treatment effects. Section~2 develops the framework, Section~3 presents the research design, Section~4 reports the correspondence and calibration results, and Section~5 discusses implications for experimental discovery in IR.

\section{Silicon Samples, Language-Based Experiments, and Behavioral Correspondence}

\subsection{Evaluating Silicon Subjects as a Convenience Sample}
Human-subject experiments are powerful, costly, and difficult to scale across settings. Political scientists have therefore used cheaper and more accessible subject pools. Online labor markets, especially Amazon Mechanical Turk, provide the clearest precedent. Their scientific use depended on evidence that treatment effects corresponded to those observed in established human samples \citep{berinsky2012mturk,mullinix2015generalizability,Coppock_2019_GeneralizingMTurk}.

This literature focuses on sample validity: whether altered sample composition changes the treatment effects used for inference. Convenience samples often recover similar treatment effects despite descriptive differences from probability samples or nationally representative benchmarks \citep{Coppock_2019_GeneralizingMTurk,CoppockLeeperMullinix_2018_HTE}. Silicon sampling raises the same evaluative question. Its scientific value depends on recovery of the inferential relationships of interest, with surface similarity serving as a secondary criterion.

The external-validity literature also examines transport across experimental environments \citep{CoppockGreen_2015_LabField}. \citet{EgamiHartman_2023_ExternalValidity} organize generalizability around changes in populations and contexts. Recent IR work applies this logic across countries, where institutions, social experience, and information environments may alter treatment-response relationships \citep{BassanNygateRenshonWeeksWeiss_2024_Generalizability}. This literature yields two questions for silicon subjects: whether the prompted sample approximates the target population and whether the experimental relationship travels into the LLM response context.

Research on silicon samples speaks most directly to the first question: how closely model responses resemble those of the target population. As descriptive simulators, LLMs remain fragile. They can generate plausible survey responses while misrepresenting response distributions, subgroup variation, and statistical relationships in human data \citep{bisbee2024synthetic,wang2025flatten,XieEtAl_2026_StatisticalRealism}. These limits are especially consequential for descriptive inference, where the object of interest is the level and distribution of human attitudes.

Models perform well on theory-of-mind benchmarks, demonstrating task success without establishing human-like mental-state reasoning \citep{StrachanEtAl_2024_ToM}. They can predict scientific results in some domains and forecast treatment effects in large collections of social-science experiments \citep{LuoEtAl_2025_NeurosciencePrediction,ashokkumar2024predicting}. GPT-style models also show structured choice patterns in economic decision tasks, with lower heterogeneity than human subjects and sensitivity to framing \citep{ChenLiuShanZhong_2023_EconomicRationality}. Related work finds content effects in model reasoning, indicating that semantic cues systematically shape outputs \citep{LampinenEtAl_2024_ContentEffects}. These findings suggest that LLMs may preserve intervention-sensitive changes in judgment even when their population-level simulations remain imperfect.

A human convenience sample can differ from the target population yet yield a similar estimate of the treatment effect \citep{Coppock_2019_GeneralizingMTurk}. This suggests a criterion for evaluating silicon responses even when their estimates are biased. If the responses contain information about the human treatment effect, giving them some weight may improve an estimate based on a small human sample. Prediction-powered inference lets us examine that possibility directly. Cross-prediction and mixed-subject designs extend this logic to model-generated observations with imperfect accuracy \citep{angelopoulos2023ppi,zrnic2024crossppi,broska2025mixedsubjects}.

Foreign-policy experiments offer a way to ask whether LLM responses to a manipulation help researchers anticipate its effect in a human experiment.

\subsection{Simulating Group Experiments in Foreign Policy}

Experiments in international relations identify the microfoundations of foreign-policy judgment by varying politically meaningful cues in controlled settings \citep{Tomz_2007_AudienceCosts,TomzWeeks_2013_DemocraticPeace,Wallace_2013_Torture,ChiltonMilnerTingley_2020_FDI,KertzerRenshonYarhiMilo_2021_Resolve,BrutgerKertzer_2018_Reputation,Kertzer_2022_Hawkish}. Researchers can organize these experiments across a design space to compare candidate treatments and scope conditions \citep{HainmuellerHopkinsYamamoto_2014_Conjoint,GrimmerMessingWestwood_2017_HTE,OfferWestortCoppockGreen_2021_Adaptive}.

LLM-based simulation can extend this experimental toolkit. Models can enter the same informational environments as human respondents and generate candidate treatment-response patterns across scenarios and group configurations at lower marginal cost \citep{manning2024automated,horton2023homo,grossmann2023ai,ziems2024large,argyle2023outofone,bisbee2024synthetic}.

Extending these simulations to advisory groups requires specifying whose views a leader hears and who makes the final decision, both features of foreign-policy institutions \citep{JostKertzerMinSchub_2024_Advisers,Saunders_2017_NoSubstitute,Saunders_2022_Elites}. Group experiments let researchers vary these rules and observe how members share information and aggregate judgments \citep{Roth_1986_LabExperiments,StasserTitus_1985_Unshared,KerrTindale_2004_GroupDecision}. In a deliberation experiment, \citet{Karpowitz_2012_Gender} found that the effect of group gender composition on participation depended on the voting rule. In a foreign-policy experiment, groups initially judged an adversary more resolved than individuals did and revised their judgments less in response to new information \citep{Wayne_2024_Diplomacy}. Group discussion can also leave the size of a treatment effect largely unchanged. In the three hawkish-bias experiments we study, neither peer deliberation nor consultation with advisers significantly attenuated the biases observed among individuals \citep{Kertzer_2022_Hawkish}.

Experiments with interacting language models give a reason to evaluate the group as a unit. Agents developed shared naming conventions through repeated local exchanges \citep{Ashery_2025_Emergent}. In simulations of political discussion, agents formed ties with others who held similar views, and opinions polarized as the exchanges continued \citep{Piao_2026_Polarization}. A direct comparison on a reasoning task also found that human and silicon groups more often improved on their average member's initial answer than on their best member's answer, although silicon groups converged earlier \citep{Tak_2026_Assembly}. Our advisory experiments address a related question in small groups whose membership and decision rules are fixed. How closely do simulated groups reproduce the treatment effects observed in human groups, and how much disagreement remains visible in members' final reports?

\subsection{Formulating Behavioral Correspondence}\label{sec:standards}
Evaluating LLM outputs as experimental evidence requires more than asking whether a model response looks plausible. In the most common use case, researchers give a language model a scenario, a treatment assignment, and sometimes a persona profile sampled from a broad population pool, then treat the model's response as a prediction of how a human subject would answer. The central question is therefore whether model-generated responses preserve the inferential relationships that researchers would have estimated from human subjects.

We formalize the problem by adapting the potential-outcomes notation used in the external-validity literature \citep{EgamiHartman_2023_ExternalValidity}. Let $i$ index a human unit, $T_i \in \{0,1\}$ denote treatment assignment, and $c \in \{H,L\}$ denote the response context, where $H$ is the human-subject context and $L$ is the LLM-mediated context. Let $Y_i(T=t,c)$ be the potential outcome for unit $i$ if assigned treatment $t$ in context $c$. The human benchmark average treatment effect is
\[
\tau_H = E_P\!\left[Y_i(T=1,H)-Y_i(T=0,H)\right],
\]
where $P$ denotes the human sample or target population. The corresponding silicon estimand is
\[
\tau_L = E_{\tilde{P}}\!\left[Y_i(T=1,L)-Y_i(T=0,L)\right],
\]
where $\tilde{P}$ denotes the prompted silicon sample. Behavioral correspondence asks whether $\tau_L$ preserves the conclusion implied by $\tau_H$, and, in stronger versions, whether the two estimands are close in magnitude. The average treatment effect is the leading case; the same comparison extends to other inferential targets, such as treatment-by-covariate interactions and the response distributions themselves.

This notation identifies two external-validity problems. The first is \textit{sample correspondence}: whether $\tilde{P}$ approximates the relevant human distribution $P$. Persona prompting supplies observed characteristics such as age, gender, ideology, education, and personality. Unobserved traits and joint distributions can still differ from those of human respondents, shifting synthetic estimates through sample composition. Distributional and covariate-conditioned diagnostics evaluate this dimension by comparing response distributions and covariate-outcome relationships.

The second problem is \textit{contextual correspondence}: whether the treatment-response relationship remains comparable when the same experiment moves from $H$ to $L$. Human respondents process treatments through cognition, emotion, memory, incentives, identity, and social experience. LLM-generated agents produce answers through model training, prompt instructions, decoding behavior, and simulated persona information. For a common respondent profile, contextual correspondence requires the model to approximate the human treatment effect:
\[
Y_i(T=1,L)-Y_i(T=0,L) \approx Y_i(T=1,H)-Y_i(T=0,H),
\]
at least for the estimand and population relevant to the research claim.

Sample correspondence places a demanding requirement on the joint distribution of traits, beliefs, and response styles. Language models often underrepresent social groups even when persona attributes are supplied \citep{bisbee2024synthetic,wang2025flatten,XieEtAl_2026_StatisticalRealism}. Contextual correspondence may be more attainable because the scenario text directly supplies the controlled intervention. This asymmetry motivates use of LLM outputs as experimental predictions whose credibility varies by estimand.

Before fielding a group experiment, a researcher might use a simulation to decide whether a treatment is worth testing when peers deliberate or when a leader makes the final decision. The prediction is useful to the extent that it anticipates how the treatment affects human decisions in that setting \citep{BassanNygateRenshonWeeksWeiss_2024_Generalizability}. We therefore compare human and silicon treatment effects within each decision setting and examine whether the effects vary with respondent characteristics in similar ways. We also compare the distributions of responses themselves. Two samples can show a similar average effect even when their members give quite different answers.

We assess directional consistency in two ways, first by comparing the signs of the human and silicon estimates and then by checking whether the results agree when we also consider statistical significance. Finally, we use calibration to probe why a biased silicon prediction might still be useful. If it retains information about the human treatment effect, adding it to a small human sample may improve the estimate of the effect’s size.

\section{Research Design}
\subsection{Replicating Kertzer experiments}
The three tasks come from \citet{Kertzer_2022_Hawkish}, whose experiments link established cognitive biases to politically consequential forms of hawkish judgment. These stylized tasks isolate recurring interpretive problems within foreign-policy decision-making. Prospect theory captures whether losses induce greater willingness to accept risky policies. The intentionality-bias task examines whether costly adverse outcomes are interpreted as deliberate hostility, accident, or uncertainty. The reactive-devaluation task captures whether a proposal is discounted when it is associated with an adversary. Each task probes a different aspect of judgment under ambiguity.

We retain the benchmark scenarios, treatments, and outcomes as foreign-policy evaluation tasks for LLM judgment. Within each task, the intervention changes the gain/loss frame, fatalities, or proposal source. Decision structure specifies whether judgments are made individually, by horizontal groups, or by hierarchical groups. We estimate each task treatment effect within each structure and compare the human and silicon estimates. Effects of structure itself, and treatment-by-structure interactions, are separate estimands.

\begin{figure}[htbp]
    \centering
    \includegraphics[width=0.98\linewidth]{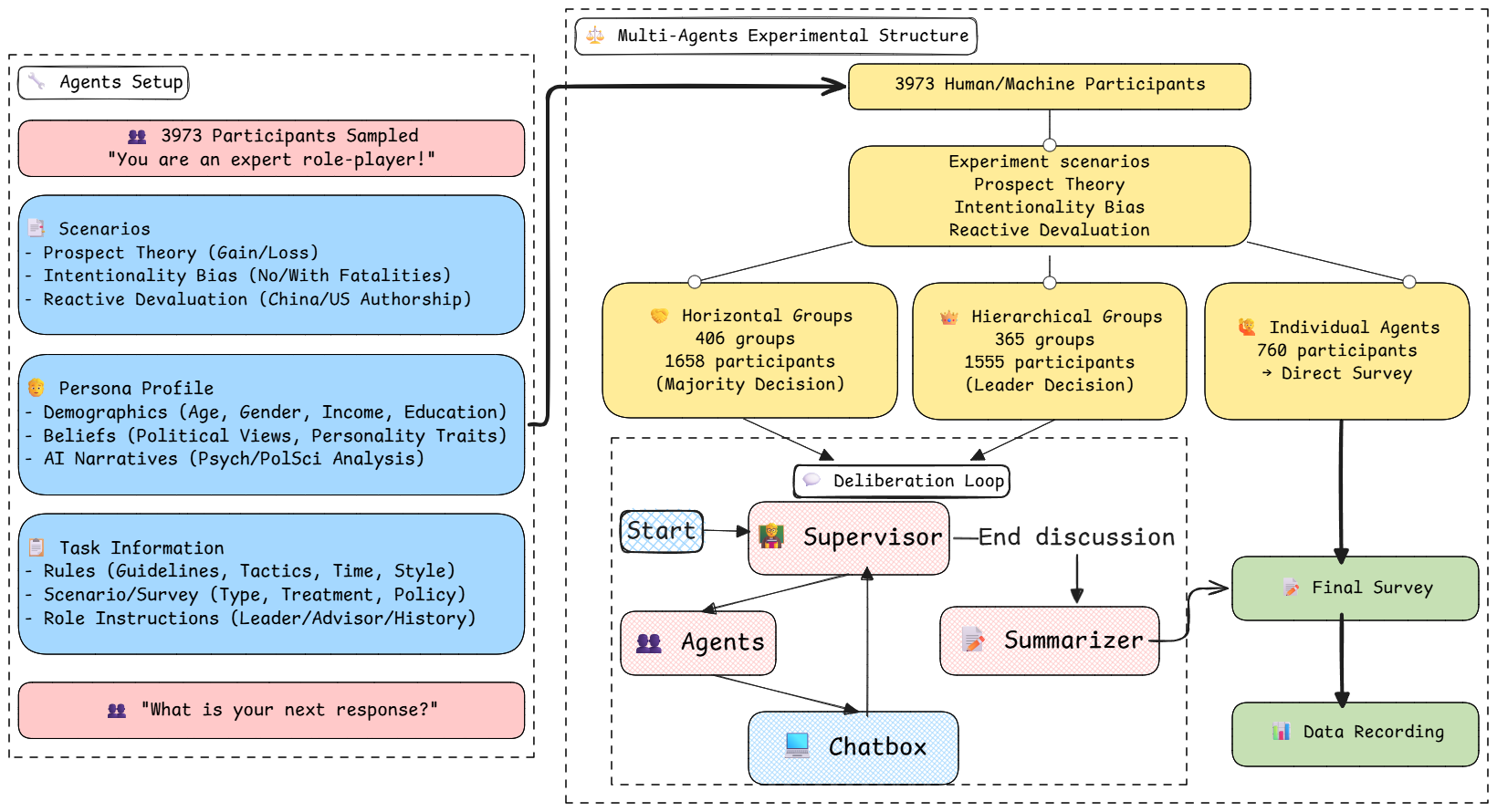}
    \caption{Experimental pipeline and decision structures}
    \label{fig:condition_split}
\end{figure}

The original benchmark also contributes an institutional design that is central to the present evaluation. Participants were assigned to one of three conditions: individual decision-making, horizontal groups, and hierarchical groups. In the individual condition, judgments were made alone. In the group conditions, participants deliberated in online chatrooms before producing decisions. Horizontal groups involved equal-status members attempting to reach a collective judgment, whereas hierarchical groups separated a leader with final authority from advisors who supplied input. These three settings allow us to compare treatment-response patterns under different arrangements for consultation and authority. Figure~\ref{fig:condition_split} shows how the implementation assigns roles, manages communication, and records individual and collective outcomes.

\subsection{Respondent Conditioning and the No-Persona Baseline}\label{sec:prompt_design}
The main individual, advisory, and heterogeneity comparisons use GPT-4o as a common model for evaluating the tested prompts and decision structures. Foreign-policy benchmarking places GPT-4o on the restrained end of model preferences, making it a conservative baseline for detecting hawkish judgment patterns \citep{jensen2025critical}. Simulation research also finds that this model class sustains differentiated personas and reproduces aggregate political judgments \citep{Park_2024_Generative,yu2024largescale}.

We first examine how respondent conditioning changes behavioral correspondence. The \textit{no-persona baseline} receives the original task text and answers directly. The \textit{primary persona configuration} receives the same task together with survey-derived question--answer pairs describing the respondent's demographic, attitudinal, and personality attributes and a narrative summary of those attributes. The model then answers directly. This construction uses respondent information to generate differentiated responses across profiles, following interview-grounded approaches to persona representation \citep{Park_2024_Generative}.

The comparison asks whether respondent conditioning brings response variation closer to the human benchmark while recovering the direction of the human treatment effects. We evaluate response dispersion and treatment effects separately. The primary configuration is also used for the subsequent individual diagnostics and is termed \textit{narrative+direct} in the calibration analyses. Appendix~\ref{app:prompt_comparison} reports the alternative prompt configurations, their comparative performance, and the rationale for selecting the primary configuration.

\subsection{Design Axes and Evaluation Dimensions}\label{sec:axis}

We evaluate predictions against the original human experiments in individual and group settings. Study 1 compares the primary persona configuration with the no-persona baseline and examines covariate-conditioned diagnostics; respondent conditioning uses existing survey profiles. Study 2 examines transport into group advisory settings. A no-persona multi-model check evaluates directional patterns without respondent profiles. Study 3 uses human calibration samples to assess bias in the predictions and tests whether adding silicon responses improves human-only estimates. Table~\ref{tab:design_summary} summarizes the tasks, studies, and correspondence targets.

\begin{table}[htbp]
\centering
\caption{Foreign-Policy Evaluation Tasks and Correspondence Checks}
\label{tab:design_summary}
\small
\begin{tabularx}{\textwidth}{p{2.5cm}Xp{4.2cm}}
\toprule
Component & Description & Correspondence target \\ \midrule
Prospect theory & Epidemic-response choice framed in terms of lives lost or lives saved. & Foreign-policy judgment task \\
Intentionality bias & Assessment of hostile intent after an adverse military event. & Foreign-policy judgment task \\
Reactive devaluation & Evaluation of the same proposal under different authorship cues. & Foreign-policy judgment task \\
\midrule
Study 1 & Individual responses with and without respondent conditioning, plus covariate-conditioned outcomes and treatment heterogeneity. & Contextual and sample correspondence \\
Study 2 & Persona-conditioned horizontal and hierarchical advisory settings. & Contextual correspondence \\
Study 3 & Calibration tests what silicon predictions contribute to a human estimate. & Calibrated utility \\
\midrule
Robustness & No-persona responses across multiple frontier models. & Cross-model directional stability \\
\bottomrule
\end{tabularx}
\end{table}

The design varies along four axes: prompt design, decision structure, correspondence dimension, and human augmentation. The main prompt comparison evaluates respondent conditioning against the no-persona baseline described in Section~\ref{sec:prompt_design}; alternative configurations appear in Appendix~\ref{app:prompt_comparison}. Decision structure covers isolated individuals, horizontal groups, and hierarchical groups. The correspondence dimension follows the framework in Section~\ref{sec:standards}: contextual correspondence covers treatments, heterogeneity, and advisory settings, while sample correspondence covers distributions and covariate-outcome relationships. Human augmentation evaluates calibrated utility. Separating these axes prevents success on one criterion from being generalized to the others.

We operationalize two directional consistency standards. The first is \textit{exact three-state correspondence}. Each estimand (treatment effect, covariate coefficient, or interaction term) is classified as a positive signal, negative signal, or non-signal using 95\% confidence intervals or $p<.05$ for regression coefficients. A cell counts as correspondence when the human and silicon classifications match exactly. Shared non-signals capture agreement in inferential classification and are reported separately from active same-direction signals when interpreting heterogeneity.

The second standard is \textit{directional sign match}, which records whether human and silicon point estimates share a sign across differences in precision. Same-sign estimates identify candidates for subsequent human fielding. Exact three-state correspondence provides the stricter inferential comparison, while directional sign match supports broad discovery at the cost of additional false positives.

\subsection{Study 1: Individual Correspondence Benchmarks}\label{sec:study1}
Study 1 collects the individual-level benchmarks, holding the foreign-policy tasks fixed and the decision structure individual. It proceeds in two layers. The first layer compares the primary persona configuration with the no-persona baseline described in Section~\ref{sec:prompt_design}, asking whether respondent conditioning preserves the individual treatment-effect conclusion while generating response variation that better matches the human benchmark across binary and ordinal outcomes.

The second layer extends the same individual setting to the conditioned diagnostics that probe whether correspondence remains visible under covariate conditioning. We examine two kinds of conditioned patterns. First, outcome-covariate correspondence compares whether persona and covariate characteristics predict outcomes in corresponding ways for human and silicon subjects. Second, treatment-heterogeneity correspondence examines whether treatment-by-covariate interaction patterns show corresponding signs across human and silicon subjects.
The two conditioned checks demand different information from the model. Treatment heterogeneity combines the controlled manipulation with a persona-based moderator, so it depends on contextual response and respondent conditioning. Outcome-covariate correspondence depends entirely on how respondent characteristics map onto baseline responses. These informational demands motivate the expectation that silicon subjects may reproduce average treatment effects more readily than relationships that depend on respondent characteristics. We assess active signals and shared non-signals separately when evaluating that expectation. Additional diversity analyses and selected transcript summaries appear in Appendix~\ref{app:diversity} and Appendix~\ref{app:transcripts}.

\subsection{Study 2: Simulating Advisory Groups}
Study 2 places the individual judge into horizontal and hierarchical advisory settings. The human benchmark comprises 760 individual respondents, 406 horizontal groups (1,658 participants), and 365 hierarchical groups (1,555 participants). The simulations retain the benchmark group identifiers, member rosters, roles, and task-specific treatment assignments. Horizontal groups deliberate as peers, while hierarchical groups assign one leader with final authority and several advisors who provide counsel before post-deliberation responses are recorded. Study 2 evaluates whether individual-level judgment patterns travel through consultation, hierarchy, and aggregation. Its target is correspondence in experimental response patterns within simulated advisory structures.

We built the group experiment in LangGraph, with one GPT-4o agent for each participant. Each agent receives the assigned version of the vignette, instructions for its role, and a persona drawn from the benchmark survey. The persona is sampled separately from the roster identifier. The coordinator advances the rounds. At the start of each round, members see the same public chat history and decide in parallel whether to contribute or pass. New messages become visible in the next round. When discussion ends, each agent privately records a decision and gives a reason. We use the median response as the peer group's decision and the leader's response as the hierarchical group's decision. Retaining each member's answer also lets us measure disagreement and check alternative rules for aggregating peer decisions. Appendix~\ref{app:methodology} describes the discussion protocol and stopping rules. Appendix~\ref{app:persona_prompting} details the prompts, and Appendix~\ref{app:implementation} explains how the two decision rules are implemented. Table~\ref{tab:prompt_context_coverage} distinguishes the group protocol from the individual prompting conditions.

The no-persona multi-model exercise supplies a robustness check for the discovery studies. Fourteen frontier models receive the original benchmark task text without persona profiles or deliberative structure. The model set spans capability tiers and laboratories in the United States and China. This appendix evidence asks whether minimally prompted models recover human treatment-effect directions across model families. Its scope is the no-persona task response; the persona and advisory pipeline is evaluated with GPT-4o. Differences across the tested models describe their performance on these tasks, while attributing those differences to training or alignment would require a design that isolates those features.

\subsection{Study 3: Assessing Silicon Predictions Through Calibration}

One explanation for the correspondence in the group experiments is that silicon subjects respond to some of the same treatment cues as humans, even when their judgments differ in other respects. We test an observable implication of this account. If silicon responses retain information about human treatment effects, giving those responses some weight can improve an estimate based on a small human sample. We calibrate individual and group estimates against matched human outcomes and examine the incremental contribution over human-only estimation in the individual experiments. We draw on the prediction-powered inference (PPI) framework developed in \citet{angelopoulos2023ppi} and extended in \citet{zrnic2024crossppi} and \citet{broska2025mixedsubjects}. The core idea is that machine-generated predictions can be combined with a smaller labeled sample to produce corrected estimands whose coverage guarantees hold even when the machine predictions are biased. In our setting, the large unlabeled sample consists of silicon-subject responses, and the smaller labeled sample consists of matched human observations. The PPI logic provides a principled basis for estimating how much human data is needed to correct silicon bias at a given level of precision.

Let $\hat{Y}$ denote the LLM-generated outcome and $Y$ the human outcome for a calibration subset. For any target estimand $\theta$, the calibrated estimator is
\[
\theta_{\mathrm{cal}}=\theta(\hat{Y}_{\mathrm{all}})+\{\theta(Y_{\mathrm{cal}})-\theta(\hat{Y}_{\mathrm{cal}})\}.
\]
The first term uses all silicon observations to estimate the quantity of interest at scale. The second term estimates the human--silicon bias in a small matched calibration sample and subtracts it from the synthetic estimate. This procedure corrects estimands (means, average treatment effects, character coefficients, and treatment-by-character interactions) without requiring a corrected individual-level dataset.

Because the full PPI correction can over-adjust when the synthetic signal is weakly aligned with human treatment-effect variation, we also consider fixed shrinkage weights $\lambda\in[0,1]$ and weights estimated from the calibration sample (Appendix~\ref{app:shrinkage_ppi}). Median estimated weights in the narrative+direct comparison are small ($\hat\lambda\approx 0.11$--$0.12$); we also evaluate fixed shrinkage, including $\lambda=0.25$.

We implement the calibration using repeated resampling across calibration shares up to 75\% of the available human results. For individual outcomes and individual average treatment effects, calibration is performed within task. For group-level outcomes, human and silicon observations are paired by task, treatment, group condition, and group identifier; hierarchical groups use the leader's decision and horizontal groups use the median aggregation rule. For character effects and treatment heterogeneity, the calibrated coefficients come from models that include standardized demographic, attitudinal, and personality traits, with treatment-by-character interactions used for heterogeneity.

The calibration exercise evaluates inferential classification and estimation error. Calibrated sign correspondence uses the exact three-state standard defined in Section~\ref{sec:standards}. Absolute error measures the distance from the full human benchmark. The ATE summaries below average errors across tasks on their recorded outcome scales and distinguish means from medians over resampling draws. Supplemental normalized diagnostics divide errors by the task outcome scale and standardize character variables to one-SD units. These measures describe how calibration performance changes with the available human sample.

\section{Evaluating the Correspondence}

\subsection{Individual Correspondence}\label{sec:directional}
This subsection reports the individual-level benchmarks of Study 1 in two layers: first the comparison of treatment effects and response dispersion with and without respondent conditioning, then the covariate-conditioned diagnostics.

\begin{figure}[htbp]
    \centering
    \includegraphics[width=\textwidth]{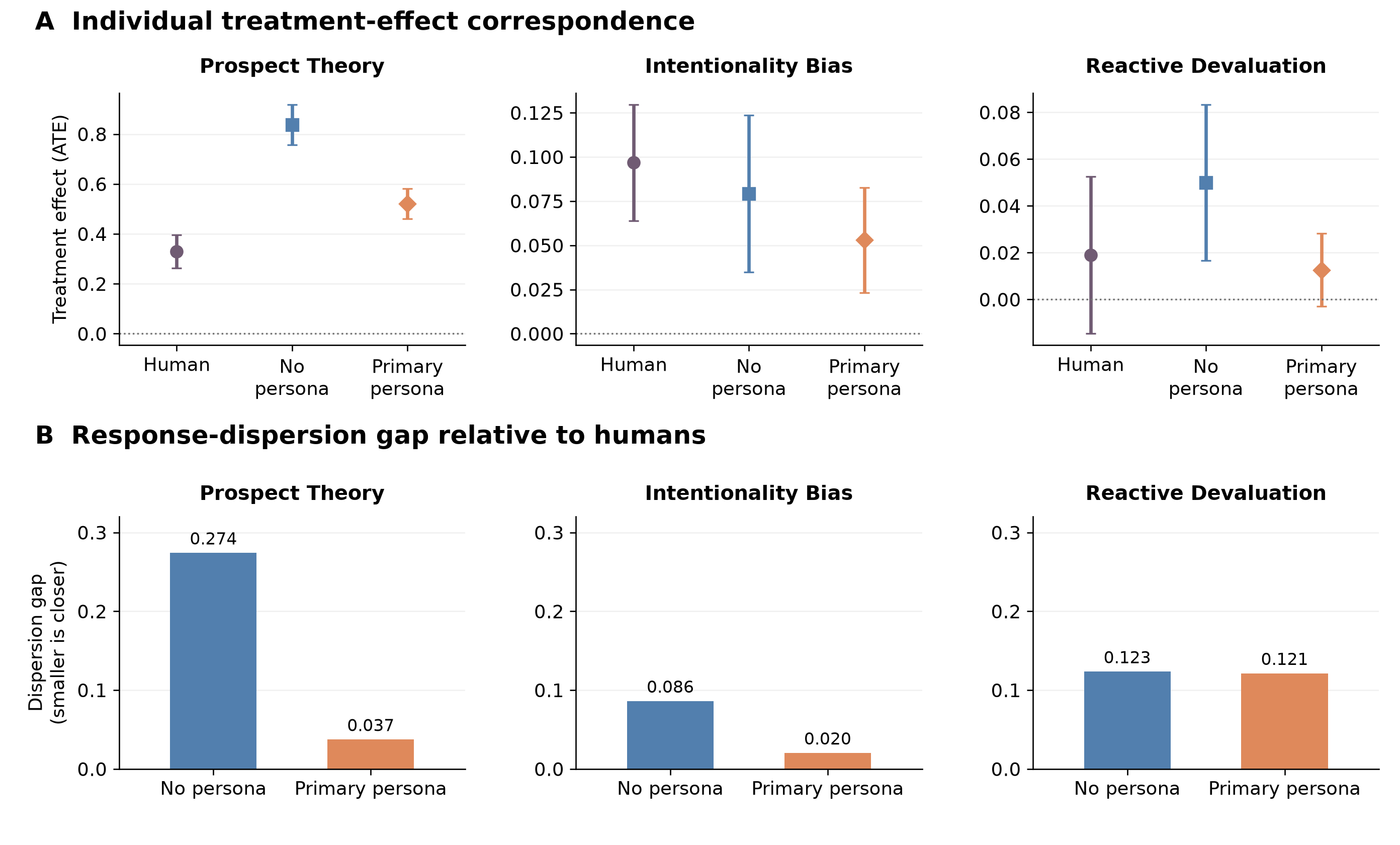}
    \caption{Respondent conditioning and individual correspondence}
    \label{fig:prompting_technique}

    \begin{minipage}{\textwidth}
        \small Note: Panel~A compares average treatment effects for human respondents, the no-persona GPT-4o baseline, and the primary GPT-4o persona configuration (QA + narrative, direct answering). Bars report 95\% confidence intervals. Effects use the outcome coding of the full prompt comparison in Appendix~\ref{app:prompt_comparison}. Panel~B reports the response-dispersion gap relative to the human benchmark; smaller values indicate closer correspondence in dispersion. Exact three-state correspondence distinguishes positive signals, negative signals, and non-signals.
    \end{minipage}
\end{figure}

Figure~\ref{fig:prompting_technique} reports the comparison against the human benchmark. Panel~A compares treatment effects for human respondents, the no-persona baseline, and the primary persona configuration. Under the exact three-state standard, the primary configuration matches all three task-level classifications: positive signals for prospect theory and intentionality bias, and a shared non-signal for reactive devaluation. The no-persona baseline matches the first two classifications and produces a positive signal for reactive devaluation.

Panel~B compares the gap between silicon and human response dispersion. Respondent conditioning reduces this gap substantially in prospect theory and intentionality bias, with little change in reactive devaluation. These results show that persona information can bring response variation closer to the human benchmark while preserving the treatment-effect classifications. Correspondence in dispersion is one dimension of distributional correspondence; Appendix~\ref{app:prompt_comparison} reports the broader distributional diagnostics and task-dependent tradeoffs across prompt configurations.

The comparison supports evaluating respondent conditioning through both sample and contextual correspondence. Appendix Figure~\ref{fig:app_no_persona_multimodel} extends the no-persona evidence across model families.

The second layer extends the individual setting to covariate-conditioned diagnostics, examining treatment-by-covariate interactions and covariate-outcome relationships separately.

Appendix Figure~\ref{fig:app_treatment_heterogeneity_correspondence} reports the treatment-heterogeneity check using the primary narrative+direct configuration. Under the exact three-state standard, 91 of 108 treatment-by-covariate cells match (84.3\%). All 91 matches are shared non-signals. The remaining cells comprise six human-only signals and eleven silicon-only signals, with zero active same-direction matches. The headline rate therefore describes agreement in nonsignificant classifications; recovery of active treatment heterogeneity remains unsupported by these cells.

Appendix Figure~\ref{fig:app_outcome_covariate_correspondence} reports the complementary outcome-covariate check under the same configuration. Exact three-state correspondence is 70.6\%. This percentage and the treatment-heterogeneity rate summarize different sets of coefficients, each with its own mix of signals and non-signals. Their numerical ordering alone cannot establish a hierarchy of substantive recovery. The evidence for discovery is clearest for predicting the direction of task-level treatment effects. Patterns that depend on respondent characteristics require separate validation.

\subsection{Contextual Correspondence Under Advisory Settings}\label{sec:structural}
\begin{figure}[htbp]
    \centering
    \includegraphics[width=\textwidth]{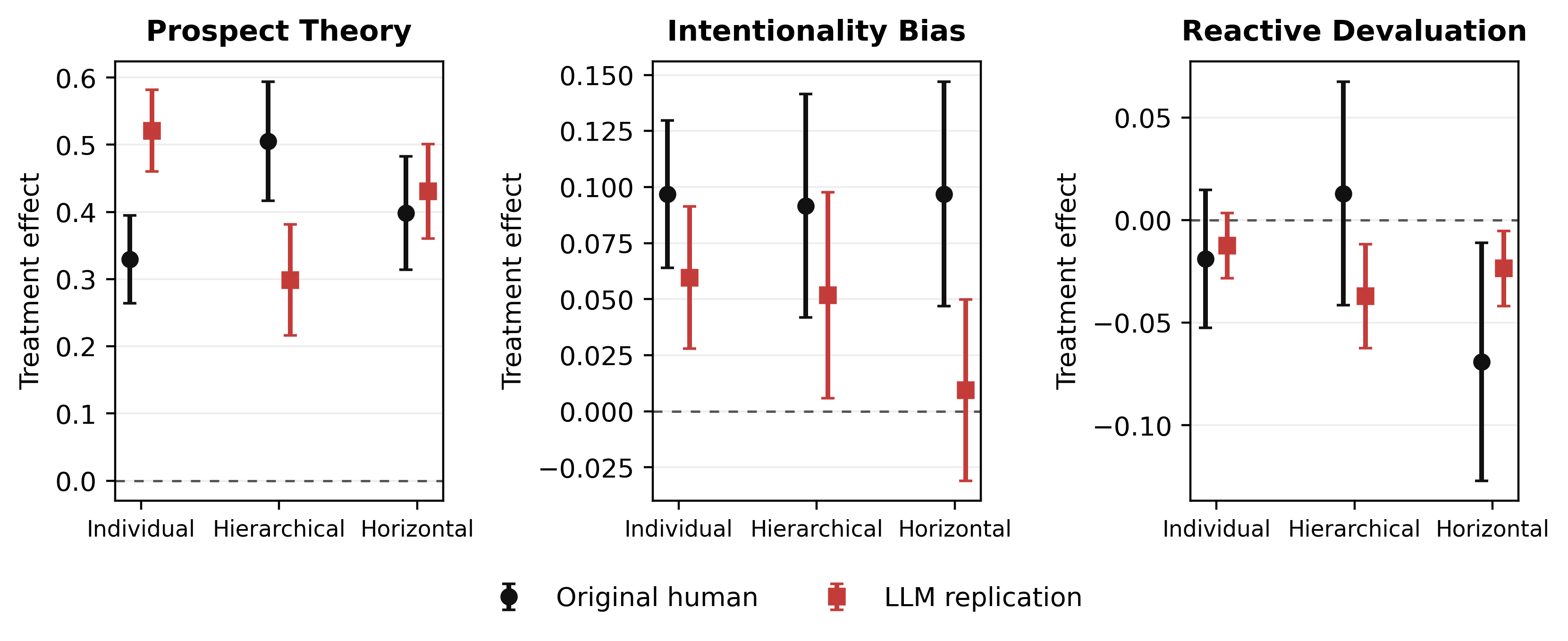}
    \caption{Human-silicon treatment effects across individual and group settings}
    \label{fig:advisory_merged}
    \begin{minipage}{\textwidth}
        \small
        \textbf{Note:} Black points and bars show the human benchmark; red points and bars show the persona-based LLM results. Horizontal group decisions follow the median-voter aggregation rule. Point estimates represent within-setting task average treatment effects, with 95\% confidence intervals. Differences across settings are descriptive comparisons of these estimates.
    \end{minipage}
\end{figure}

The persona-based advisory settings test contextual correspondence across role prompts, interaction, and hierarchy. Figure~\ref{fig:advisory_merged} compares human and silicon treatment effects across individual, hierarchical, and horizontal conditions. Correspondence varies by task and institutional structure.

Prospect theory is the clearest carryover case: the loss frame increases risky choice among human subjects and GPT-4o-based agents in all three conditions. Intentionality bias aligns in the individual and hierarchical settings; in the horizontal setting, the human effect is a positive signal and the silicon effect is a non-signal. Reactive devaluation aligns in the individual and horizontal settings. In the hierarchical setting, the point estimates have opposite signs, with a nonsignificant human estimate and a negative silicon signal. This pattern describes correspondence within each structure; testing whether the effect differs across structures requires a treatment-by-structure interaction test.

Appendix Figure~\ref{fig:app_group_treatment_correspondence} summarizes the pooled group-level results under the two standards in Section~\ref{sec:standards}. Using the paper-faithful hawkish-coded advisory cells and 95\% intervals, exact three-state correspondence is 66.7\% and directional sign match is 83.3\%. Four of six cells preserve the inferential category, and five preserve direction. The hierarchical reactive-devaluation cell illustrates the distinction between classifications and coefficient differences: despite its opposite point-estimate signs, the human--silicon coefficient-difference test has $p=.089$. Appendix~\ref{app:correspondence_diagnostics} reports the coefficient-test interpretation alongside the classification summaries.

The loss frame increased risky choice in both human and silicon advisory groups. Among matched hierarchical groups, both populations showed disagreement on this task. The mean within-group variance of final choices was 0.142 for human groups and 0.178 for silicon groups. For intentionality bias and reactive devaluation, silicon members' reports were more tightly clustered. Each task showed the same pattern in horizontal groups, although human respondents who perceived unanimity reported the group's position. Appendix Table~\ref{tab:matched_dissensus} gives the paired estimates and sample coverage.

The selected transcript summaries in Appendix~\ref{app:deliberation} and Appendix~\ref{app:transcripts} illustrate possible explanations for these task differences. Prospect-theory agents anchor on persona-consistent risk preferences and defend them through deliberation. Intentionality-bias agents emphasize the phrase ``without further information'' and converge on information-seeking caution, giving fatalities less weight as a cue of hostile intent. In reactive devaluation, hierarchical leaders preserve prior positions while advisors reshape the proposal's enforcement logic. These examples suggest how textual cues, role instructions, and aggregation may contribute to the observed group patterns.

\subsection{Calibration Results}\label{sec:ppi_calibration}
\begin{figure}[htbp]
    \centering
    \includegraphics[width=0.95\linewidth]{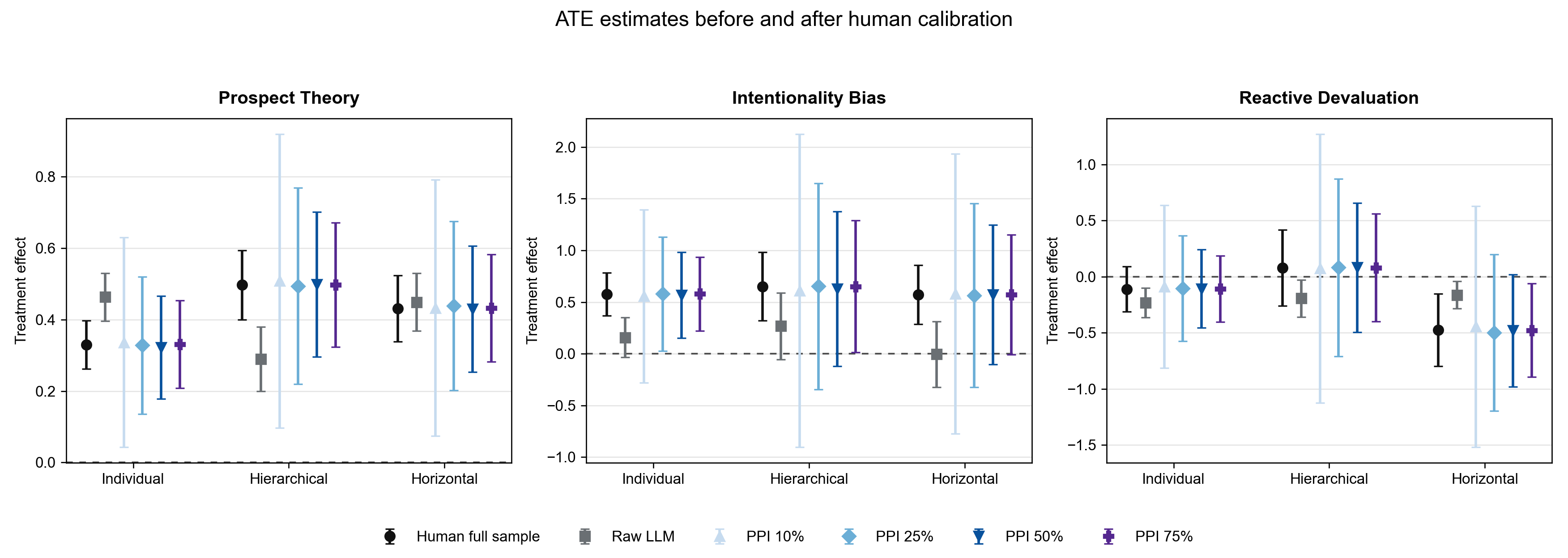}
    \caption{Average treatment effects before and after human calibration}
    \label{fig:ppi_ate_effects}

    \begin{minipage}{0.95\linewidth}
        \small Note: Points report treatment-effect estimates; intervals report 95\% normal-approximation intervals. Black circles mark the full human benchmark; gray squares mark the raw LLM estimate; blue markers show PPI-calibrated estimates at 10\%, 25\%, 50\%, and 75\% human calibration shares. Individual estimates use narrative+direct predictions. Group estimates use the advisory protocol, with hierarchical leader outcomes and horizontal median outcomes.
    \end{minipage}
\end{figure}

We assess the analogy with a biased but informative human sample using two comparisons. Improvement over the raw silicon estimate shows what human calibration corrects. Improvement over an estimate from the same human observations alone shows what the silicon predictions contribute. Figure~\ref{fig:ppi_ate_effects} compares the full human benchmark, the raw LLM estimate, and calibrated estimates at 10\%, 25\%, 50\%, and 75\% human calibration shares. Individual estimates use narrative+direct predictions; group estimates use responses from the persona-based advisory protocol.

Calibration moves point estimates toward the human benchmark. For individual treatment effects, the average of the three task-level median absolute errors is 0.112 at 25\% human labels, compared with 0.224 for the raw LLM. It falls to 0.065 at 50\% and 0.038 at 75\%. These values summarize typical error across repeated calibration draws; the corresponding mean absolute error at 25\% labels is 0.135. Group-level treatment effects follow the same pattern, with wider intervals in advisory settings. These comparisons establish correction relative to raw LLM estimates. For the individual experiments, Appendix~\ref{app:shrinkage_ppi} compares calibration methods with equally sized human-only samples. At 25\% labels, mean absolute error in that comparison is approximately 0.085 for estimated-weight PPI, 0.086 for human-only estimation, and 0.100 for full PPI. The small gain from estimated weights is distinct from the larger correction of raw LLM error.

\begin{figure}[htbp]
    \centering
    \includegraphics[width=0.75\linewidth]{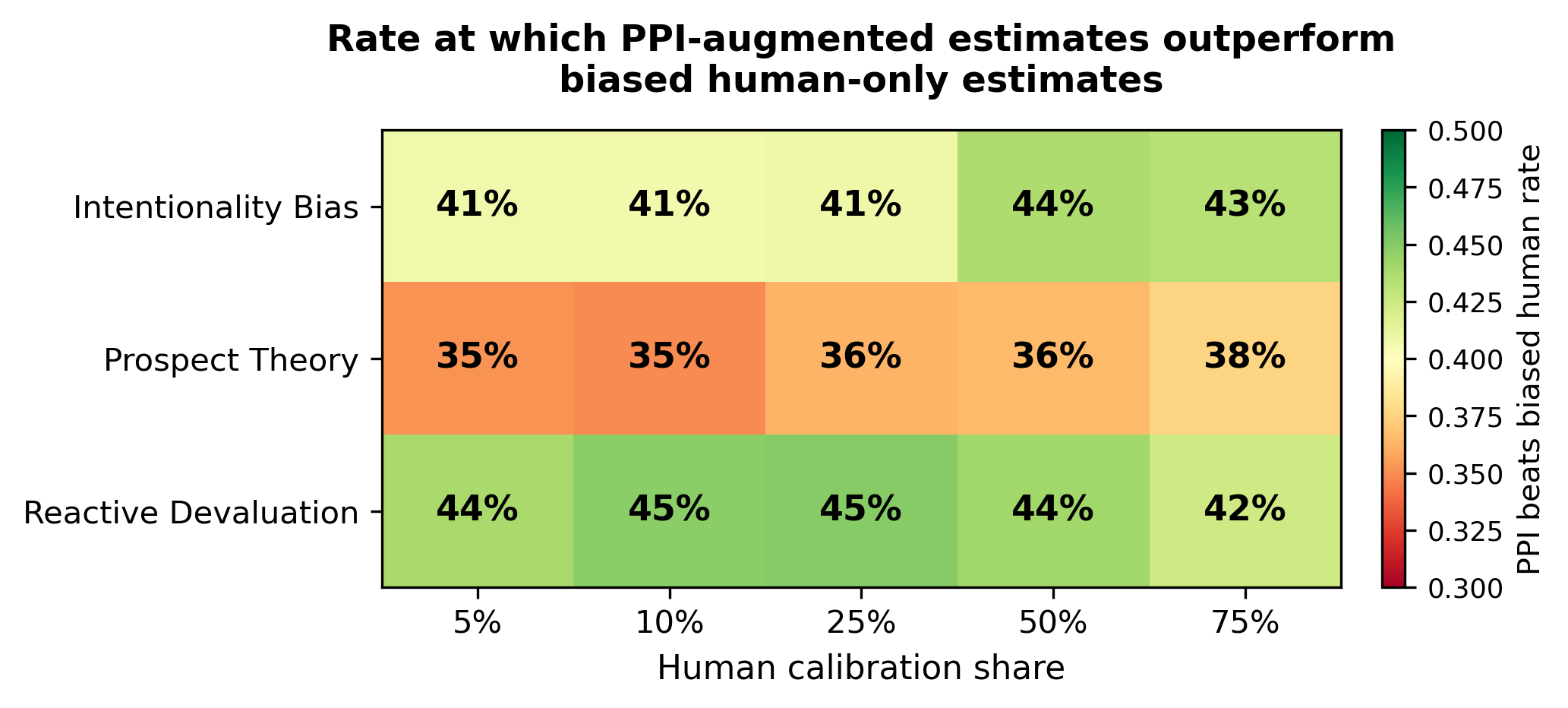}
    \caption{PPI win rate against biased human-only estimates}
    \label{fig:biased_sample_comparison}

    \begin{minipage}{0.95\linewidth}
        \small Note: Each cell reports the share of simulation draws in which PPI-calibrated ATE error is smaller than biased human-only ATE error, averaged across 16 bias scenarios and 500 draws per cell. All estimates use the narrative+direct LLM configuration. The full bias-scenario breakdown is reported in Appendix~\ref{app:biased_sample}.
    \end{minipage}
\end{figure}

In the individual experiments, we ask what silicon responses can add when the human sample overrepresents some respondent profiles. We draw human calibration samples under 16 compositional biases across eight covariates, at shares from 5\% to 75\%. In each draw, PPI starts with the treatment effect for the sampled humans. It adds the difference between the silicon treatment effect for all matched profiles and the silicon effect for the sampled profiles. Across 120,000 draws, this adjustment pointed toward the full human benchmark in 53.9\% of draws and reduced absolute error in 40.7\% (Figure~\ref{fig:biased_sample_comparison} reports the latter comparison by task and calibration share). An adjustment in the right direction could still overshoot. Mean absolute error was higher with PPI in each task-by-share cell.

Human labels sharply correct raw LLM treatment-effect estimates. Against an equally sized human-only sample, feasible shrinkage PPI yields small gains in balanced samples. Under the tested compositional biases, silicon predictions improve estimates in a substantial minority of draws, while human-only estimates have lower average error. Appendix~\ref{app:shrinkage_ppi} extends the comparison to covariate coefficients and treatment heterogeneity, where calibration requirements are higher.

\section{Discussion and Conclusion}

This article evaluates silicon subjects through \textit{behavioral correspondence}: the degree to which their outputs reproduce specified experimental estimands observed among human subjects. The framework separates contextual correspondence in treatments, heterogeneity, and advisory structures from sample correspondence in distributions and covariate-outcome relationships. The practical question is which predicted treatment effects warrant testing in a human experiment.

Silicon subjects often capture the direction of human treatment effects, which makes them useful for screening treatments before fielding a group experiment. Simulated and human groups can respond similarly to a treatment even when their members disagree to different degrees. Predicting which respondents are most affected or how a group reaches its decision requires separate evidence. Calibration suggests why a biased silicon sample can still be useful. Like a human convenience sample, it can carry information about a treatment effect even when it misestimates its size. That information sometimes improved estimates from a small human sample, although under the sampling biases we tested the human-only estimates were more accurate on average.

Future research could extend the advisory experiments of \citet{Kertzer_2022_Hawkish} to a polarized issue on which the same evidence may persuade some advisers and harden the views of others. Researchers could compare which advisers change their judgments and how their groups reach a decision, probing why human and silicon groups sometimes respond in the same direction \citep{FuSlough_2026_HTE_Mechanisms,Fu_2026_ExtractingMechanisms}. Structured literature reviews and retrieval-augmented generation could draw on prior studies to build scenarios and respondent profiles for these experiments. Human outcomes would then provide a benchmark for testing whether those additions improve predictions of who responds to a treatment and whether the predictions add information to estimates from a small human sample.

\newpage
\bibliographystyle{apsr}
\bibliography{refs}

\clearpage  

\appendix
\setcounter{page}{1}
\setcounter{table}{0}
\setcounter{figure}{0}
\setcounter{equation}{0}
\setcounter{footnote}{0}
\renewcommand\thetable{A\arabic{table}}
\renewcommand\thefigure{A\arabic{figure}}
\renewcommand{\thepage}{A-\arabic{page}}
\renewcommand{\theequation}{A\arabic{equation}}
\renewcommand{\thefootnote}{A\arabic{footnote}}

\onehalfspacing
\section{Supplementary Materials}
\label{app:supplement}

\bigskip
{\bf\small
\begin{itemize}\itemsep0.45em
\item[A.1.] The Methodology of Synthetic Replication
\begin{itemize}\itemsep0.35em
\item[A.1.1.] The General Architecture and Environment
\item[A.1.2.] The High-Level Pseudo-Code
\item[A.1.3.] Long-Term and Short-Term Memory
\item[A.1.4.] Agentic Reasoning: Fast and Slow Thinking
\end{itemize}
\item[A.2.] Prompt Engineering and Synthetic Persona Construction
\item[A.3.] Synthetic Replication Implementation and Efficiency
\item[A.4.] Robustness of the Synthetic Replication: Correspondence Assessment
\item[A.5.] Sign Correspondence Diagnostics
\item[A.6.] Coefficient Plots for Interpretability Diagnostics
\item[A.7.] Analyses of Diversity and Deliberation 
\begin{itemize}\itemsep0.35em
\item[A.7.1.] Does Diversity Affect Group Decision-Making?
\item[A.7.2.] Disagreement in Reported Final Decisions
\item[A.7.3.] Qualitative Analyses of Agent Deliberation
\end{itemize}
\item[A.8.] Selected Deliberation Transcript Summaries
\begin{itemize}\itemsep0.35em
\item[A.8.1.] Prospect Theory, Gain Frame (Horizontal Group)
\item[A.8.2.] Intentionality Bias, All Fatalities (Horizontal Group)
\item[A.8.3.] Reactive Devaluation, China Authorship (Hierarchical Group)
\end{itemize}
\item[A.9.] Shrinkage PPI and Estimated-$\lambda$ Diagnostics
\item[A.10.] Biased-Sample Calibration Test
\item[A.11.] LLM Configuration Comparison
\end{itemize}
}
\clearpage

\subsection{The Methodology of Synthetic Replication}
\label{app:methodology}
\subsubsection{The General Architecture and Environment}
To establish a robust, fully automated solution for LLM-based laboratory experiments, we constructed a custom multi-agent environment using LangGraph, a framework designed for building complex agent-based applications. We selected this framework for three core advantages: its graph-based philosophy, which allows complex deliberation workflows to be visualized and executed as directed cycles; its robust state management, which facilitates easy tracing and logging of every interaction; and its integrated debugging system, which ensures stability during large-scale simulations. Consequently, our entire experiment is organized as a cyclical workflow loop flowing from a central coordinator to parallel agent participants, then to a chatbox processing unit, and finally to a private survey system, all wrapped in comprehensive error logging mechanisms.

At the center of this architecture is a structured shared discussion state. This centralized memory persists across the entire simulation and is logically divided into input configurations, temporary execution data, and output results. It contains the public chat history, agent settings (long-term memory including persona profiles and roles), agent statuses (short-term tracking of current actions), and scenario-level information. Crucially, it also maintains a single global time variable that tracks the simulated time elapsed in the conversation. This state object updates as it moves through the system, ensuring that every component, from the coordinator to the individual agents, has access to the same global environment.

The environment is initialized via a dedicated module before any agents speak. In this step, the system loads the specific experimental scenario (e.g., the gain versus loss frame in the Prospect Theory module), assigns the group structure (Horizontal or Hierarchical), and attaches the relevant synthetic persona profiles to each agent.

A critical design choice in our replication was the interaction logic. Instead of a simple sequential turn-taking format, we implemented a simultaneous update mode to mimic the dynamics of a real-world online chatroom. In this configuration, all agent participants execute in parallel within a given step. They review the existing history and decide whether to speak based on the information visible at that moment. Their outputs are then sent to a processing unit, which aggregates the messages, sorts them by their simulated completion timestamps, and reformats the history. The resulting discussion allows participants to formulate thoughts simultaneously instead of waiting for a rigid order. A timing function calculates simulated "subjective" time from reading, thinking, and speaking, thereby constraining agent participation.

To ensure the process requires zero manual intervention, we implemented a dedicated Coordinator component. This unit manages the progression of rounds and includes a deliberation monitoring function. This monitor continuously evaluates the discussion state, detecting specific signals from the agents, such as consensus achievement or explicit "end" markers, to autonomously conclude the deliberation loop.

Finally, the architecture includes a distinct Survey component. Once the coordinator detects the end of the deliberation phase, the system transitions to a survey mode. In this phase, agents are isolated to provide private, persona-consistent responses to post-discussion questionnaires, so that each member reports a final judgment without seeing the others' survey answers.

\subsubsection{The High-Level Pseudo-Code}
Figures~\ref{alg:pipeline_a}–\ref{alg:pipeline_b} report the full implementation as high-level pseudo-code, from persona calibration to LangGraph orchestration and estimation of experimental quantities. The full code, configuration files, and raw outputs are provided in the replication materials.

\begin{figure}[htbp]
    \centering
    \begin{minipage}{0.97\linewidth}
\begin{verbatim}
Algorithm 1 (part 1/2): Replicating the advisory groups

Input:
    benchmark survey, original group records, task vignettes, model

1. Set up each recorded advisory group
   - Read its group ID, structure, member IDs, roles, and treatment
     assignment for each task from the benchmark.
   - Create one agent per roster entry and retain its ID and role.
   - Separately sample a survey respondent's answers for the persona;
     use its question-answer pairs and narrative in the agent prompt.
2. Environment, memory, prompts, and discussion graph
   - Define DiscussionState with fields:
       chatbox, agent_settings, agent_statuses, scenario,
       current_round, objective_time, agent_responses, results.
   - Long-term memory: store persona QA pairs, role (leader / advisor /
     individual), group structure, scenario metadata in agent_settings.
   - Short-term memory: track chatbox by round, time stamps, token-based
     time metrics, and member status (including overtime).
   - Define an agent-prompt template combining:
       role description, persona profile, scenario + treatment,
       discussion history, and meta-instructions on interaction strategy.
   - Record each model reply, reasoning, and token counts.
   - Build the discussion and private survey routes:
       initialize -> coordinator -> agents in parallel -> chat processor
       -> coordinator for the next round or the stopping decision.
       At the stop, send each member its private survey;
       end the run after collecting the answers.
   - Attach checkpointing to persist DiscussionState across steps.
\end{verbatim}
    \end{minipage}
    \caption{High-level pseudo-code for the multi-agent replication pipeline (part 1/2): roster and persona setup and construction of the LangGraph discussion environment.}
    \label{alg:pipeline_a}
\end{figure}

\begin{figure}[htbp]
    \centering
    \begin{minipage}{0.97\linewidth}
\begin{verbatim}
Algorithm 1 (part 2/2): Replicating the advisory groups

3. Run the advisory experiments
   for each recorded group in the run:
     agents <- agents for this roster, with their recorded roles
     for each task:
       state <- initialize(task, group's assigned treatment, agents)
       while the coordinator continues discussion:
         history <- public chat at the start of this round
         for each member in parallel:
           replies[member] <- contribute or pass given history,
                              persona, role, and assigned vignette
         update chat and time after collecting the round's replies
       for each member:
         final[member] <- private survey choice and reason,
                          given the discussion and the task
       save(group ID, roster, roles, task, treatment, chat, final)
       choices <- code members' private final choices
       if peer group: primary outcome <- median(choices)
       if hierarchical group: outcome <- designated leader's choice
4. Analysis and comparison with human experiments
   - Construct an analysis dataset from logged results:
       outcomes (support levels, attribution scales, etc.),
       treatment indicators, group-level diversity indices
       (demographics, dispositions, experience, political attitudes).
   - Using the same coding as in the original study, compute:
       average treatment effects (ATEs) for humans and LLMs,
       within-group disagreement and polarization measures.
   - Compare sign and magnitude of ATEs and dissensus across species
     and group structures, and evaluate how they change with diversity.
   - Export replication-ready tables and figures used in the main text
     and in the diversity appendix.
\end{verbatim}
    \end{minipage}
    \caption{High-level pseudo-code for the multi-agent replication pipeline (part 2/2): experimental runs, aggregation rules, and construction of replication-ready estimands.}
    \label{alg:pipeline_b}
\end{figure}

\subsubsection{Long-Term and Short-Term Memory}
The multi-round interaction can lead to bloated and confusing context for LLMs. We manage the agent's context through a long- and short-term memory framework. In our framework, long-term memory enters through the agent settings that are fixed at initialization. For each agent, we specify a role (for example, leader or advisor), a concise statement of role responsibilities (for example, guiding discussion and making a final decision versus providing advice to the leader), and a personality profile. The personality profile is constructed from question–answer pairs derived from survey data or synthetic profiles and covers demographic attributes, political predispositions, ideology, religious background, personality traits, risk attitudes, and work experience. In the prompt, these QA pairs are presented as if they were the transcript of a prior interview between the agent and a researcher. This design ties the agent's behavior to a structured, interpretable representation of ``who they are" without requiring the model to infer a persona from scratch.

Short-term memory is entirely contained in the discussion state. It consists of the chatbox history, a set of agent statuses (including their most recent time metrics and any penalties), the current round, and additional flags such as whether a survey is currently being administered. Before each call to the LLM, the system builds a personalized view of the history for the focal agent, concatenates it with the persona interview and role description, and includes the current objective time and round number. Each agent therefore “remembers” its individual background and the evolving group discussion, while the framework controls which parts of the history are visible and how they are framed.

\subsubsection{Agentic Reasoning: Fast and Slow Thinking}

The framework separates ``fast" and ``slow" components of the simulated decision process. During deliberation, the model receives role and persona instructions, the discussion history, and guidance to justify major positions in a short, conversational message. The private survey requests a selected answer and its reasoning. When the API exposes reasoning through a separate channel, the framework stores it in the agent's status and hides it from other agents. When the API returns a single text output, the framework parses reasoning and message segments using simple delimiters. Prompt, reasoning, and content tokens provide approximate measures of reading, thinking, and speaking time across experimental conditions and group structures.

Agents receive concise system instructions that define their role, emphasize persona consistency, and specify interaction strategies. The core prompt instructs each agent to ``stay in character," engage through agreement and disagreement, identify perceived mistakes, act strategically, and adjust response length to the remaining time and discussion progress. Agents may say ``[pass]" when they have nothing new to add, which creates alternating participation and observation. In time-sensitive scenarios, the prompt treats thinking time as costly and links concise reasoning to additional speaking opportunities.

\subsection{Prompt Engineering and Synthetic Persona Construction}
\label{app:persona_prompting}

The prompt engineering framework follows a four-part structure. First, meta-instructions define the task as a prediction task for an expert role-player. The model is told that it is playing the role of a particular participant in a group decision-making setting, communicating via a chatbox, and that its job is to predict what this person would say next given their characteristics and the situation. Second, a synthetic persona profile anchors the agent's behavior in realistic demographic and psychological attributes. Third, task-specific information provides the scenario description, experimental manipulation, decision question, and group structure. Finally, concluding meta-instructions translate all of this into a concrete behavioral output, specifying approximate length, tone, and the need to remain consistent with the persona and past messages.

For example, the opening meta-instruction is of the form: ``You are an expert role-player, now playing as a participant in a simulated group decision-making process and communicating through a chatbox. Your persona profile is as follows: [synthetic persona profile]. The current task is: [answer a survey question or participate in discussion]. Considering your persona's core beliefs, your prior responses, and the guidelines, what is your next response?" We constrain answers to a few dozen words and require consistency with the persona's background and previous behavior.

Synthetic persona construction proceeds in several stages. We first extract calibration variables from the original experimental survey data, including demographics, party identification, ideological self-placement, religious identity, personality items, risk attitudes, and indicators of professional and political experience. These variables form the backbone of the persona. We then derive a set of question-answer pairs that reconstruct each respondent's responses as a natural-language interview. For each item, the question is rendered in plain language and the response is translated into a short textual answer that preserves the meaning of the underlying scale. To improve interpretability and encourage consistent behavior, we additionally generate short narrative summaries of each persona that synthesize the demographic and belief attributes into a coherent description of the person's worldview and decision-making style. In the prompt, the persona appears as the transcript of an interview followed by a brief narrative, which gives the model both granular detail and a higher-level interpretive frame.

Task information is layered on top of the persona. For each experiment, we provide a complete description of the scenario and treatment, the specific policy proposal or crisis situation at issue, and the survey question that operationalizes the outcome variable (for example, which policy to choose, how likely a hostile intent is, or how much to support a proposal). We also describe the group’s decision rule and the agent’s role within that rule: individual respondents answer only for themselves; horizontal group members participate as equal advisors whose individual final answers are later aggregated; hierarchical groups have one leader and the remaining members serve as advisors, and only the leader’s final answer is treated as the group outcome. To preserve continuity, we include a short summary of the discussion history in each round so that agents can explicitly refer back to previous comments and adjust their contributions accordingly.

\subsubsection{Prompt Configurations and Selection of the Primary Configuration}
\label{app:prompt_comparison}

The individual comparison evaluates four practical prompt configurations. The no-persona baseline supplies the original task text and requests a direct answer. QA only adds a respondent profile rendered as question--answer pairs. QA + CoT adds an instruction to reason step by step before answering \citep{wei2022chain}. QA + narrative supplies the question--answer profile together with a narrative characterization of the respondent's personality, worldview, and decision style, followed by direct answering \citep{Park_2024_Generative}.

These are selected comparisons. No persona versus QA only evaluates the addition of respondent information; QA only versus QA + CoT evaluates the reasoning instruction; QA only versus QA + narrative evaluates narrative representation alongside the same profile information. QA + CoT versus QA + narrative changes both narrative representation and the answering instruction. The supplemental calibration comparison in Appendix~\ref{app:config_comparison} additionally includes narrative+CoT, covering QA/narrative representation crossed with direct/CoT answering in that analysis.

\begin{table}[htbp]
\centering
\small
\caption{Prompt configurations in the individual comparison}
\label{tab:prompt_configurations}
\begin{tabular}{llll}
\toprule
Configuration & QA profile & Narrative summary & Answering \\
\midrule
No persona & Absent & Absent & Direct \\
QA only & Present & Absent & Direct \\
QA + CoT & Present & Absent & Step-by-step \\
QA + narrative & Present & Present & Direct \\
\bottomrule
\end{tabular}
\end{table}

\begin{figure}[htbp]
    \centering
    \includegraphics[width=\textwidth]{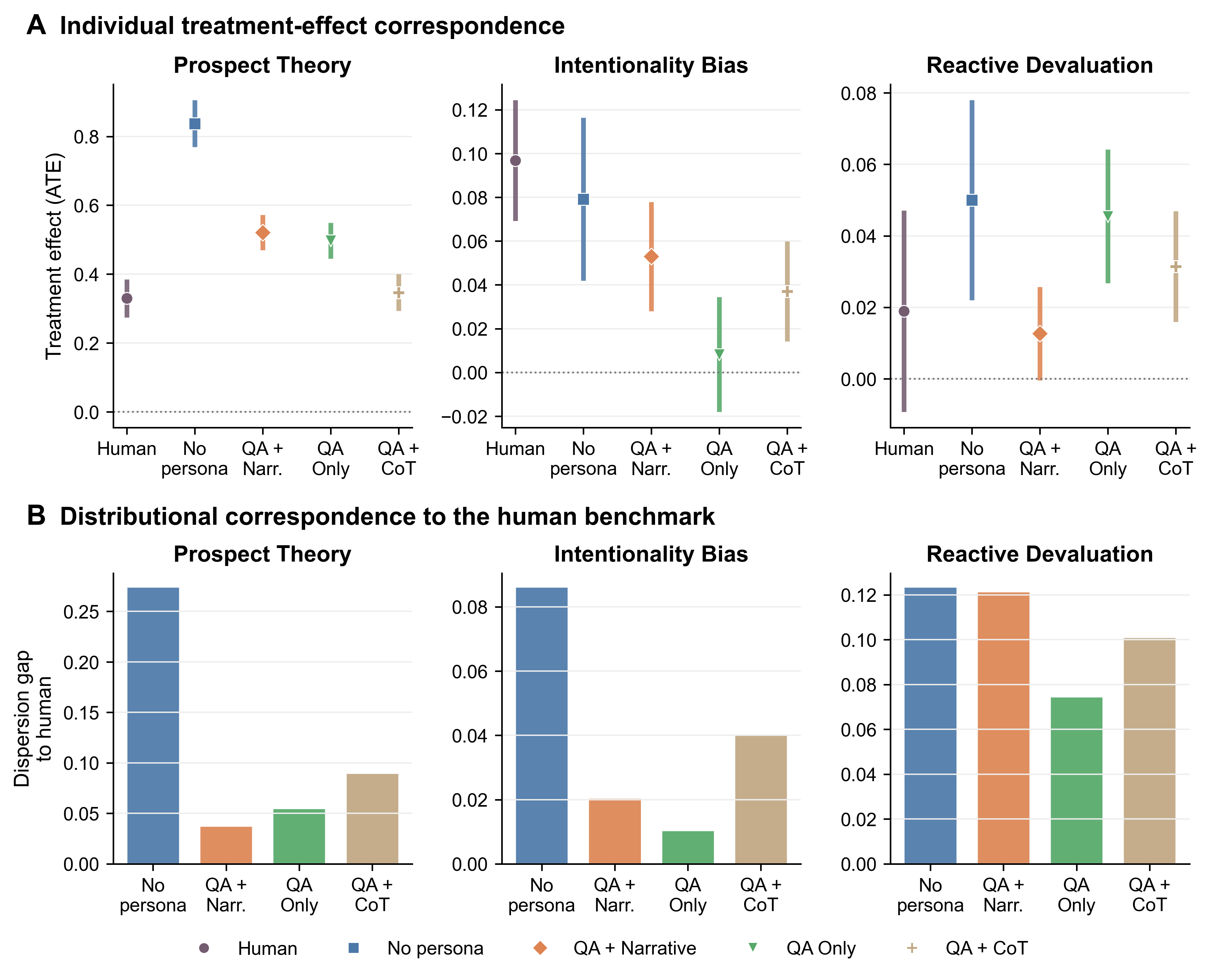}
    \caption{Full comparison of prompting techniques and individual correspondence}
    \label{fig:app_prompting_technique}
    \begin{minipage}{\textwidth}
        \small Note: Panel~A compares human treatment effects with no persona, QA only, QA + CoT, and QA + narrative. Bars display 90\% confidence intervals; the classification counts reported in the text use 95\% intervals from the corresponding estimates. Panel~B reports the response-dispersion gap relative to the human benchmark, with smaller values indicating closer correspondence in dispersion.
    \end{minipage}
\end{figure}

Figure~\ref{fig:app_prompting_technique} retains the full individual comparison. At the 95\% classification threshold used in the main text, QA + narrative matches all three human treatment-effect classifications, including the shared non-signal for reactive devaluation; QA + CoT matches two, and QA only matches one. These classifications are distinct from closeness in effect magnitude: for example, QA + CoT is closer to the human prospect-theory estimate than QA + narrative.

The distributional comparison also involves tradeoffs. Relative to no persona, QA + narrative reduces the dispersion gap from 0.274 to 0.037 for prospect theory and from 0.086 to 0.020 for intentionality bias, with little change for reactive devaluation (0.123 to 0.121). Wasserstein distance to the human response distribution falls for prospect theory (0.256 to 0.102), but rises for intentionality bias (0.078 to 0.178) and reactive devaluation (0.249 to 0.291). Closer dispersion therefore coexists with differences in the broader response distribution.

We select QA + narrative as the primary configuration on the basis of its treatment-effect classification correspondence and improvements in response dispersion relative to no persona. This is a benchmark-based selection using the same tasks reported in the evaluation. Its performance varies by task and criterion. Appendix~\ref{app:config_comparison} separately evaluates calibrated treatment-effect error across the four persona-by-reasoning configurations.

\subsubsection{Coverage Across Prompts and Decision Structures}
\label{app:prompt_context_coverage}

Table~\ref{tab:prompt_context_coverage} distinguishes the individual answering configurations from the advisory protocol. The individual direct conditions request an answer without an added step-by-step reasoning instruction. Advisory agents exchange messages before giving private responses; the group survey instructions request an answer and its reasoning. Thus, reasoning and justification during group interaction are separate from the direct/CoT comparison among isolated respondents. The reported design permits comparison of within-setting task effects under the advisory protocol. Prompt-by-structure interactions would require the corresponding prompt configurations to be evaluated across structures.

\begin{table}[htbp]
\centering
\small
\caption{Reported prompt coverage by decision structure}
\label{tab:prompt_context_coverage}
\begin{tabularx}{\textwidth}{>{\raggedright\arraybackslash}X >{\raggedright\arraybackslash}p{2.6cm} >{\raggedright\arraybackslash}p{2cm} >{\raggedright\arraybackslash}p{2cm}}
\toprule
Protocol & Individual & Horizontal & Hierarchical \\
\midrule
No persona, direct & Study 1; 14-model check & --- & --- \\
QA only, direct & Prompt comparison; calibration & --- & --- \\
QA + CoT & Prompt comparison; calibration & --- & --- \\
QA + narrative, direct & Primary comparison; calibration & --- & --- \\
QA + narrative, CoT & Supplemental calibration & --- & --- \\
QA + narrative with deliberation and private response justification & --- & Study 2; calibration & Study 2; calibration \\
\bottomrule
\end{tabularx}
\begin{minipage}{\textwidth}
\vspace{0.5em}
\small Note: A dash indicates a combination absent from the reported comparison. The group protocol combines persona information with role instructions, discussion history, and a private final response. Horizontal outcomes use the median response; hierarchical outcomes use the leader's response. The table describes the coverage of the reported analyses; the prompt and structure factors are only partially crossed.
\end{minipage}
\end{table}

\subsection{Synthetic Replication Implementation and Efficiency}
\label{app:implementation}

In the horizontal group condition, members receive identical scenario information alongside distinct personas and deliberate as equals aiming for consensus. Each advisor then records an individual final answer. We treat the distribution of these answers as the group's outcome, with the main estimand based on the median response. In the hierarchical group condition, advisors deliberate with a single leader. Advisors provide input to the leader, who receives a reminder in each round about final decision authority. The leader's final answer after deliberation defines the group outcome. The individual condition uses the same persona and scenario prompts and omits the deliberative component.

Running these experiments with synthetic subjects is substantially cheaper than recruiting human participants while allowing us to impose much tighter control over timing, attrition, and compliance. Based on pilot runs, the average LLM API cost per group per experiment is approximately three RMB. Conducting three experiments across all 771 groups (406 horizontal and 365 hierarchical) therefore requires roughly 6,939 RMB (about USD 980 at current exchange rates). This cost includes the full deliberation process in every round and the generation of chain-of-thought reasoning. The resource efficiency makes it feasible to run extensive robustness checks and sensitivity analyses that would be prohibitively costly in conventional human-subject experiments.

\subsection{Robustness of the Synthetic Replication: Correspondence Assessment}
\label{app:correspondence_assessment}

This appendix assesses two features of the synthetic replications. The first is adherence to specified personas and roles. The second is correspondence in outcome distributions and treatment-effect directions relative to the human benchmark.

The assessment follows the main-text distinction between sample and contextual correspondence. Moving from human respondents to LLM agents changes the prompted sample and the response context. Directional correspondence records whether treatment and control differences preserve their sign in the synthetic setting. A consistent sign supports treatment-response carryover for that task and setting; a reversal or non-signal identifies a boundary for subsequent human investigation.

Prompt adherence is assessed qualitatively and quantitatively. Transcript summaries record whether agents invoke persona attributes, maintain positions across rounds, and follow assigned advisory roles. Coefficient diagnostics examine whether specified traits predict baseline preferences across experimental conditions. These checks supplement the aggregate treatment-effect comparisons.

\subsection{Sign Correspondence Diagnostics}
\label{app:correspondence_diagnostics}

This appendix collects the figures supporting the sign correspondence diagnostics in the main text. The two directional consistency standards, exact three-state correspondence and directional sign match, are defined in Section~\ref{sec:standards}. The headline correspondence rates for the treatment-heterogeneity and outcome-covariate checks are reported in Section~\ref{sec:directional}, and the group-level rates in Section~\ref{sec:structural}. Separate human--silicon coefficient-difference tests in the six group cells yield unadjusted $p<.05$ for hierarchical prospect theory ($p=.002$), horizontal intentionality bias ($p=.005$), and horizontal reactive devaluation ($p=.040$). The other three differences have $p>.05$, including hierarchical reactive devaluation ($p=.089$). These tests compare human and silicon treatment effects within each structure. Testing whether treatment effects differ across structures requires an interaction test.

\subsubsection{No-Persona Multi-Model Robustness}
The no-persona multi-model results are reported here as robustness evidence. They show whether directional treatment patterns appear across a broader frontier-model set under minimal prompting, while leaving the main empirical block focused on prompting technique and the exact three-state comparison.

\begin{figure}[htbp]
    \centering
    \includegraphics[width=0.95\linewidth]{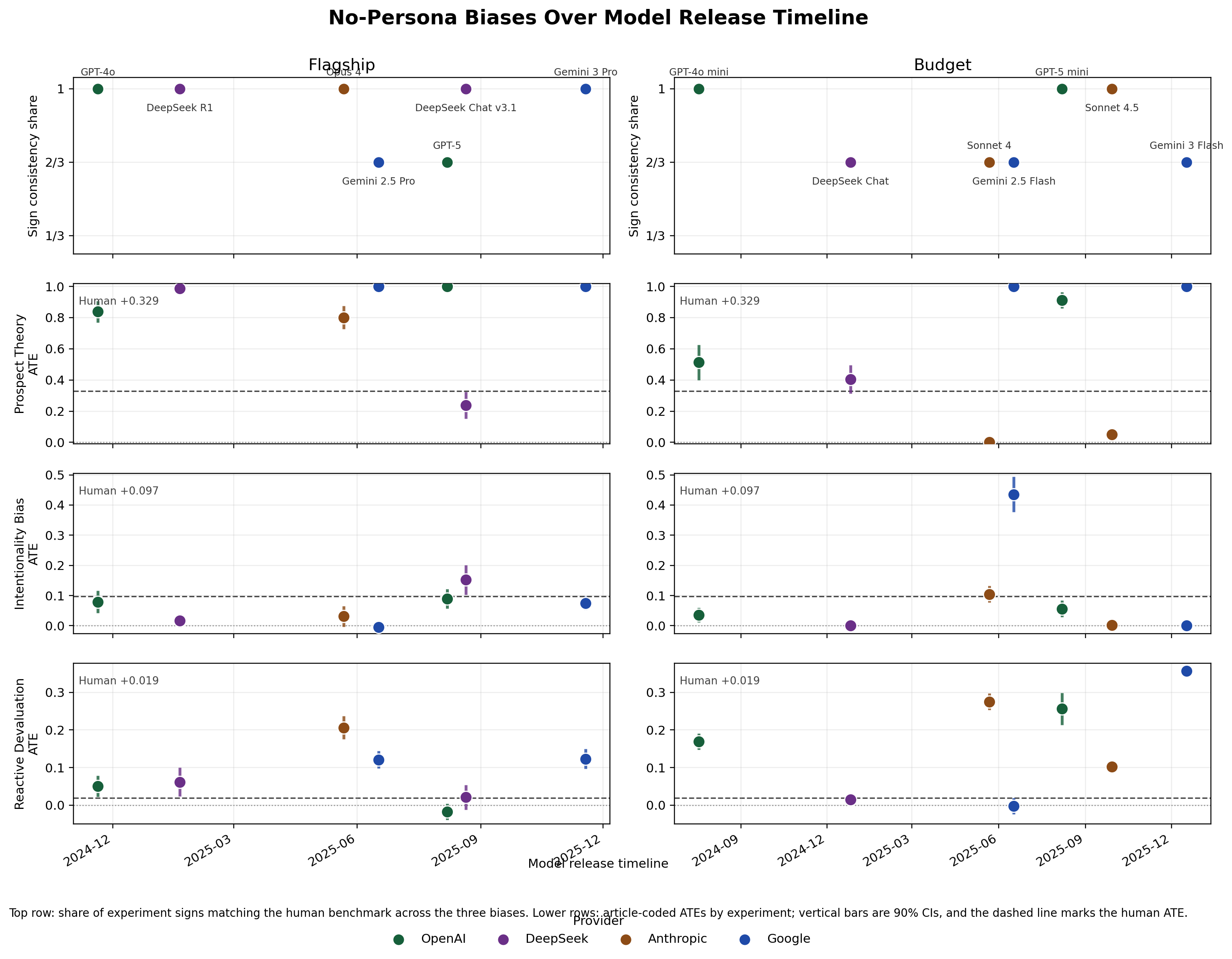}
    \caption{No-persona multi-model robustness}
    \label{fig:app_no_persona_multimodel}

    \begin{minipage}{0.95\linewidth}
        \small Note: The figure reports treatment-effect correspondence for fourteen frontier models prompted without persona profiles. The evidence provides a supplementary check of directional treatment carryover across model families; the prompt-design comparison remains the main empirical baseline. Additional models can be evaluated with the same task text, treatment coding, outcome mapping, and correspondence criteria, with model versions and decoding settings reported for each extension.
    \end{minipage}
\end{figure}

\begin{figure}[htbp]
    \centering
    \includegraphics[width=\textwidth]{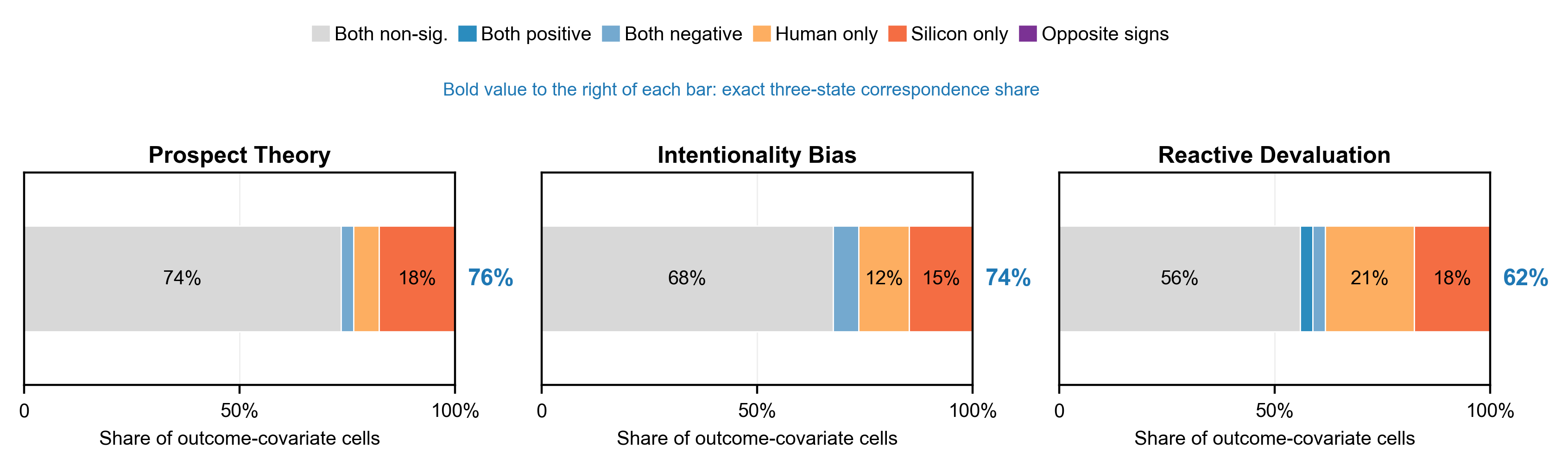}
    \caption{Outcome-covariate correspondence}
    \label{fig:app_outcome_covariate_correspondence}

    \begin{minipage}{\textwidth}
        \small Note: The figure compares whether persona and covariate characteristics predict outcomes in corresponding ways among human and silicon subjects. Each stacked bar reports the three-state classification (shared non-signal, shared positive signal, shared negative signal, human-only signal, silicon-only signal, or opposite signs) by task; the bold value to the right of each bar is the exact three-state correspondence share.
    \end{minipage}
\end{figure}

\begin{figure}[htbp]
    \centering
    \includegraphics[width=\textwidth]{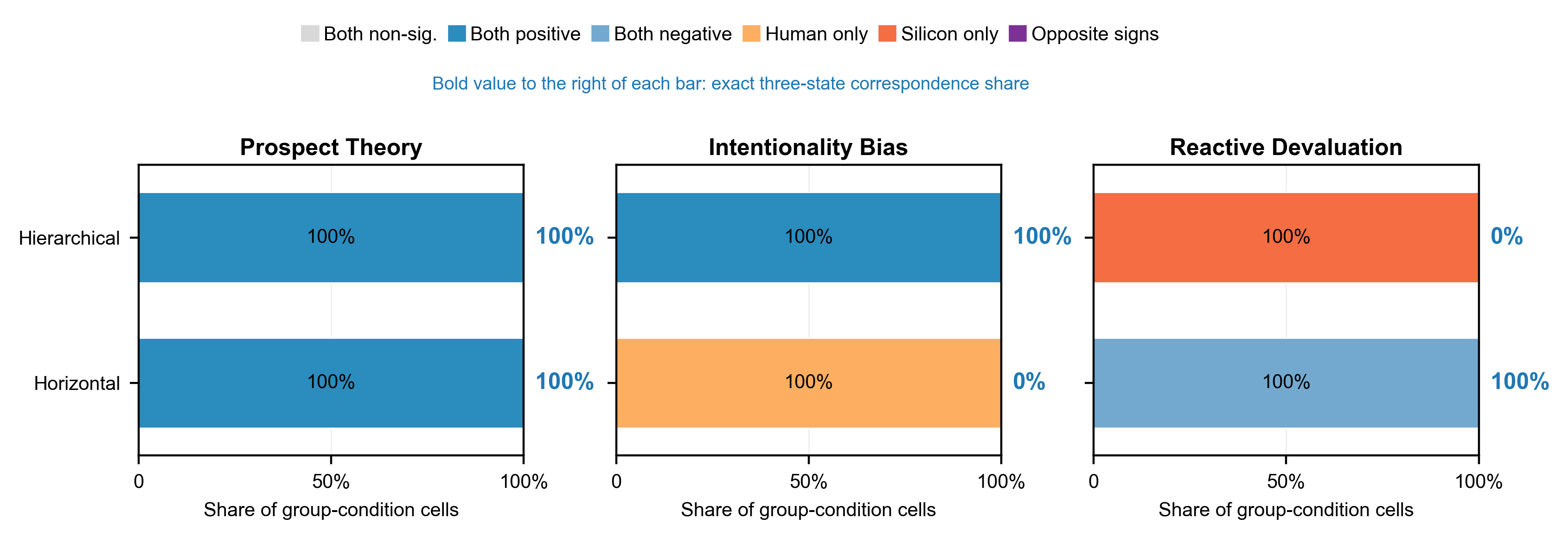}
    \caption{Group-level treatment correspondence}
    \label{fig:app_group_treatment_correspondence}

    \begin{minipage}{\textwidth}
        \small Note: Each cell is a module-by-group-condition comparison against the original human group data. Each stacked bar reports the three-state classification (shared non-signal, shared positive signal, shared negative signal, human-only signal, silicon-only signal, or opposite signs) by task and group condition; the bold value to the right of each bar is the exact three-state correspondence share.
    \end{minipage}
\end{figure}

\begin{figure}[htbp]
    \centering
    \includegraphics[width=\textwidth]{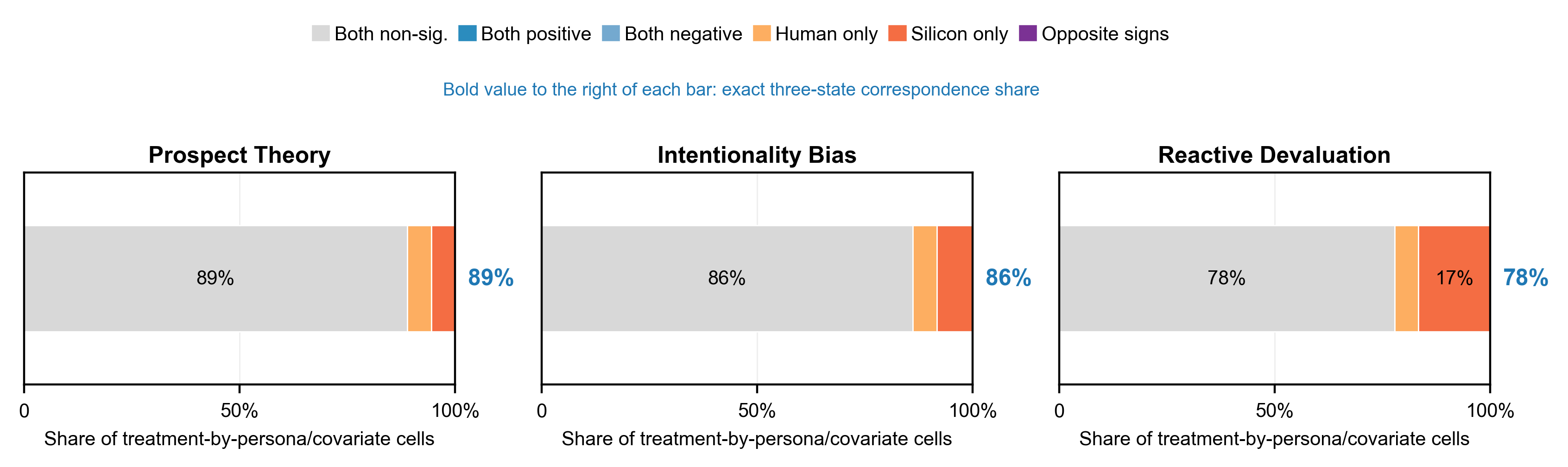}
    \caption{Treatment-heterogeneity correspondence}
    \label{fig:app_treatment_heterogeneity_correspondence}

    \begin{minipage}{\textwidth}
        \small Note: Cells compare treatment-by-persona/covariate interactions by task and AI setting. Each stacked bar reports the three-state classification (shared non-signal, shared positive signal, shared negative signal, human-only signal, silicon-only signal, or opposite signs); the bold value to the right of each bar is the exact three-state correspondence share.
    \end{minipage}
\end{figure}

\subsection{Coefficient Plots for Interpretability Diagnostics}
\label{app:coefplots}

\begin{landscape}
\begin{figure}[htbp]
    \centering
    \includegraphics[width=0.75\linewidth]{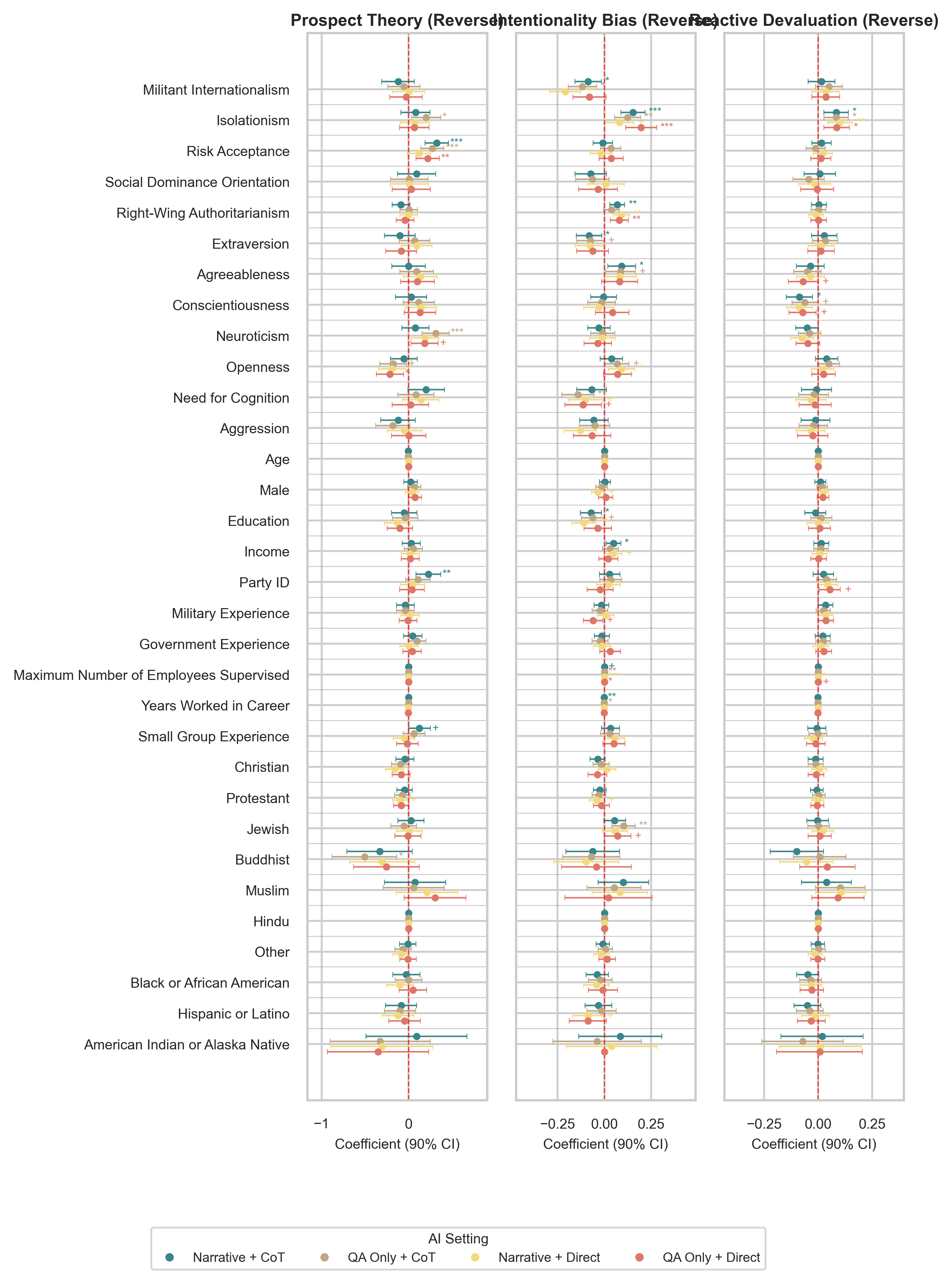}
    \caption{Persona attributes and LLM-human response distance: coefficient plot}
    \label{fig:distance_coefplot}

    \begin{minipage}{0.95\linewidth}
        \small Note: Points are regression coefficients of persona attributes on the absolute distance between LLM and human responses, with 90\% confidence intervals. Panels correspond to the three foreign-policy tasks and colors denote the four AI settings. For intentionality bias and reactive devaluation, distance dependent variables have been normalized to $[0,1]$. Explanatory variables include pretest covariates.
    \end{minipage}
\end{figure}
\end{landscape}

\begin{landscape}
\begin{figure}[htbp]
    \centering
    \includegraphics[width=0.68\linewidth]{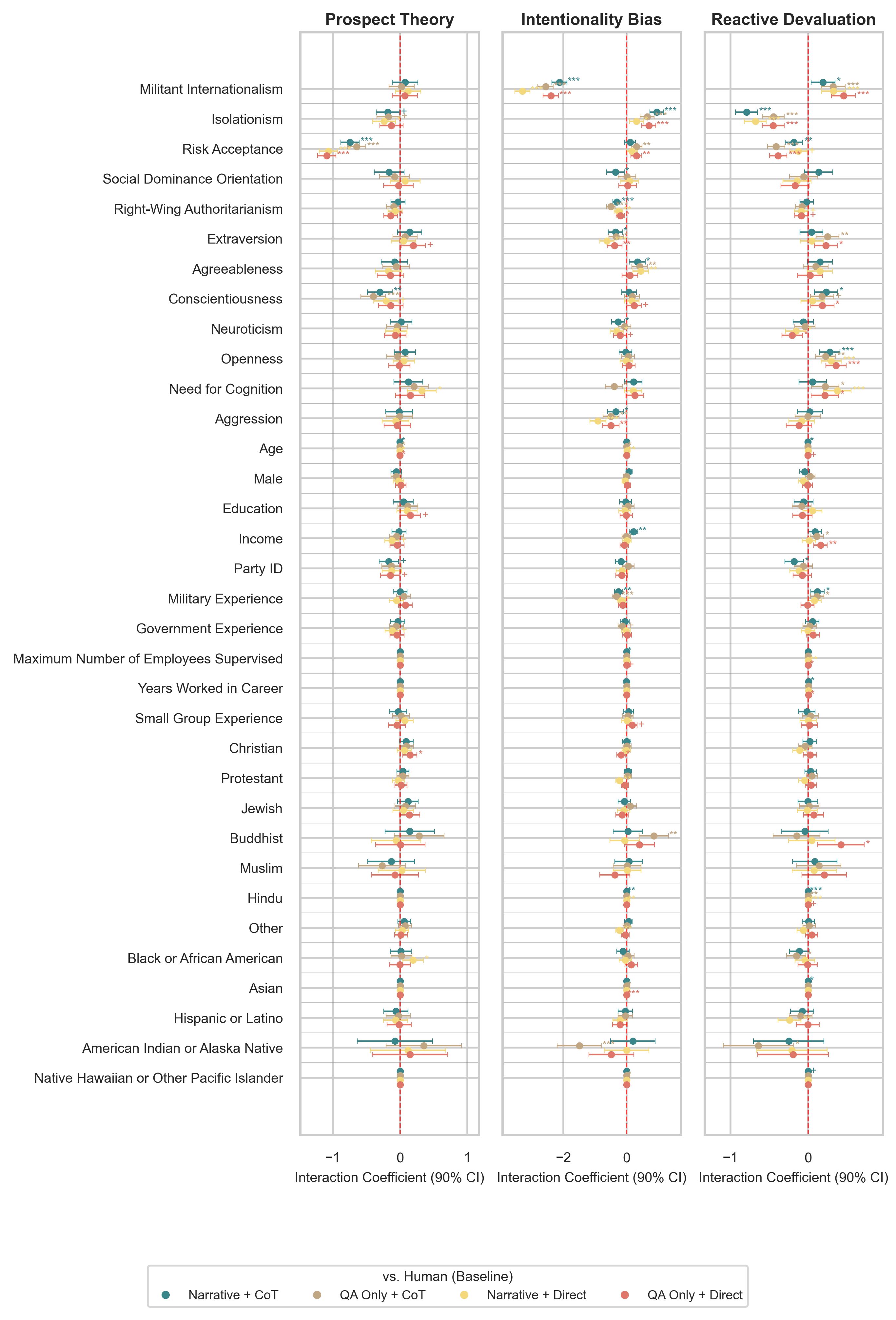}
    \caption{Persona-conditioned instability across AI settings: coefficient plot}
    \label{fig:interaction_coefplot}

    \begin{minipage}{0.95\linewidth}
        \small Note: Points are coefficients of covariate-setting interaction terms from a regression of the LLM or human response on persona attributes interacted with AI settings, with 90\% confidence intervals. Panels correspond to the three foreign-policy tasks and colors denote the four AI settings relative to the human baseline. For intentionality bias and reactive devaluation, dependent variables have been normalized to $[0,1]$. Explanatory variables include pretest covariates, settings (human as baseline and all AI settings), and their interaction terms.
    \end{minipage}
\end{figure}
\end{landscape}

This appendix reports exploratory coefficient-level diagnostics using 90\% confidence intervals. The main correspondence classifications use 95\% intervals; the exploratory plots use 90\% intervals to display coefficient patterns with greater resolution. Figure~\ref{fig:distance_coefplot} shows coefficients from the regression of LLM-human response distance on persona attributes by task and AI setting. Figure~\ref{fig:interaction_coefplot} shows interaction coefficients from a regression of responses on persona attributes interacted with AI settings, using humans as the baseline.

\subsection{Analyses of Diversity and Deliberation }
\label{app:diversity}
\subsubsection{ Does Diversity Affect Group Decision-Making?}
\begin{figure}[h!]
    \centering
    \includegraphics[width=\linewidth]{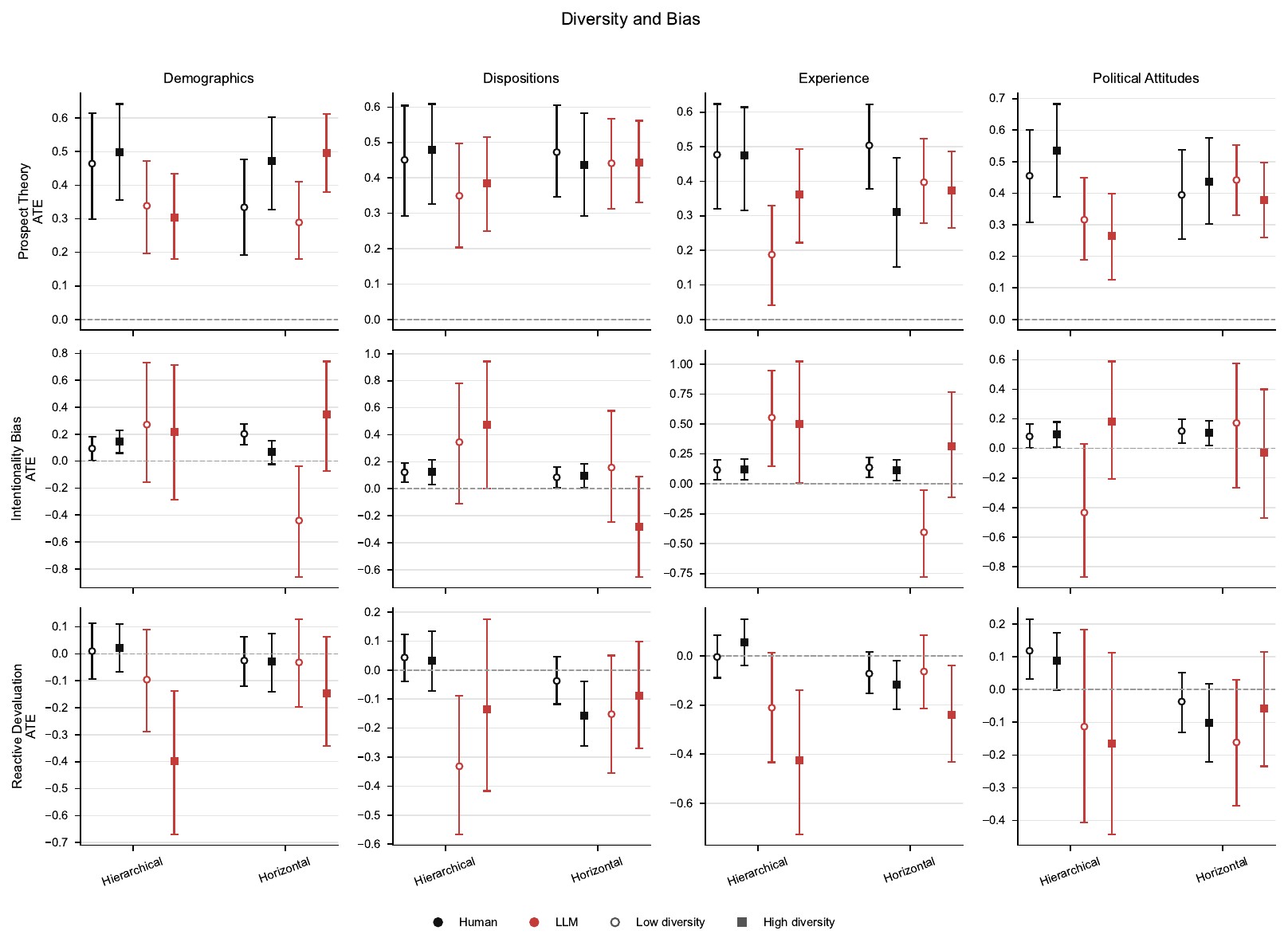}
    \caption{Diversity and Bias}
    \label{fig:F9}
\end{figure}

Following the original study, we examined whether group diversity dampens or amplifies hawkish biases in synthetic replications by comparing ATEs for less diverse groups (25th percentile and below) and more diverse groups (75th percentile and above). Figure~\ref{fig:F9} reports the results. Black points represent estimates from human subjects, while red points represent estimates from LLM subjects. Each row corresponds to one experimental module (prospect theory, intentionality bias, reactive devaluation), and each column operationalizes diversity using a different metric based on group members' demographics, dispositions, experience, or political attitudes. Point estimates are cell means with 90\% bootstrapped confidence intervals. Horizontal group decisions are calculated here using the median-voter aggregation rule.

Across all four diversity dimensions, average bias shows little systematic attenuation or amplification for human or LLM groups. Prospect-theory risk seeking remains positive across diversity levels and decision structures. Intentionality-bias estimates are weaker and less stable among LLM groups, while reactive devaluation remains larger among LLMs. Movement from low to high diversity lacks a consistent direction in either sample. Wider LLM confidence intervals indicate greater uncertainty within diversity cells.

\begin{figure}[h!]
    \centering
    \includegraphics[width=\linewidth]{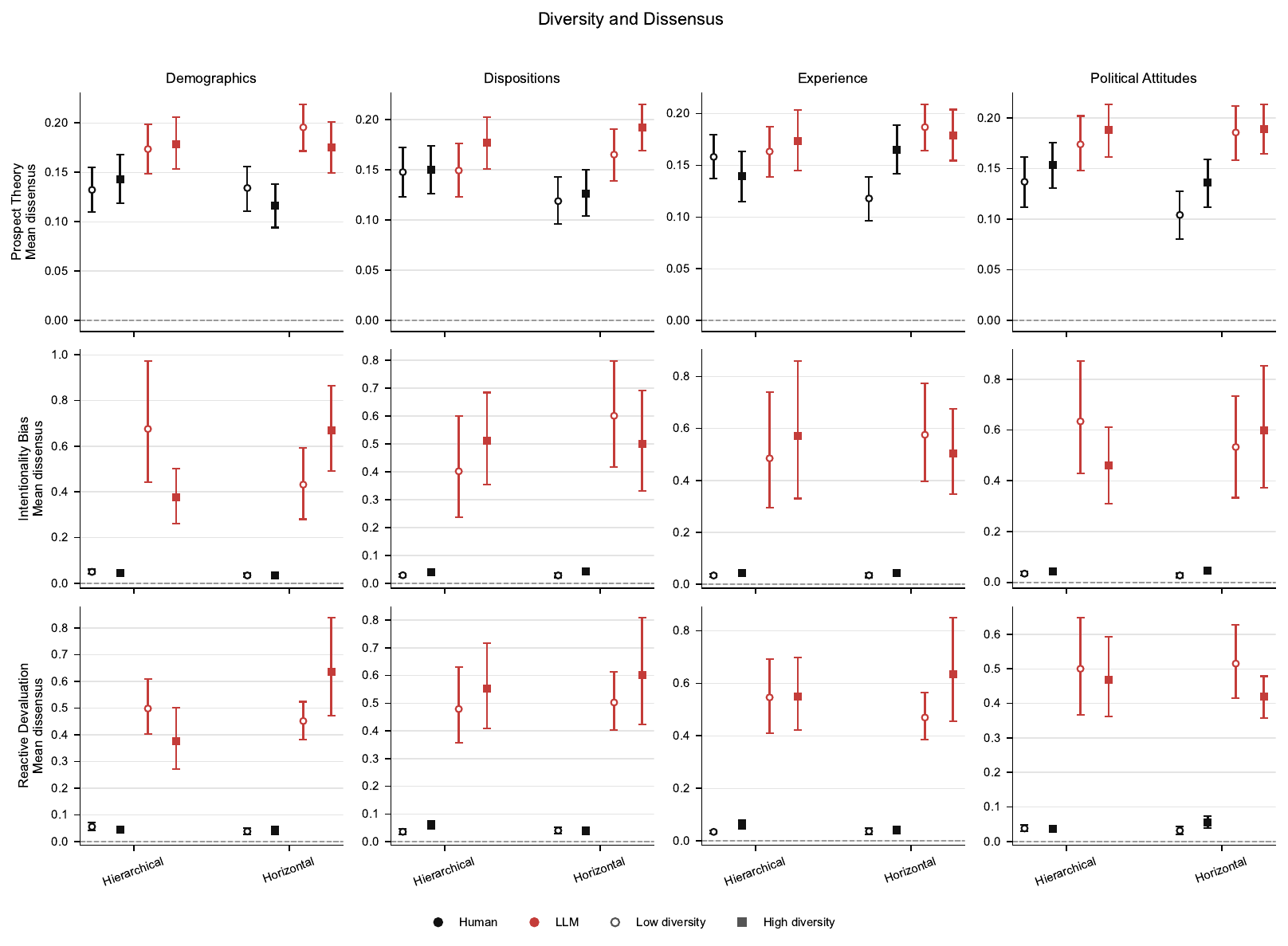}
    \caption{Diversity and Dissensus}
    \label{fig:F10}
\end{figure}

Figure~\ref{fig:F10} shifts the focus from bias magnitude to internal disagreement and plots dissensus by group diversity within each decision structure. Black points and colored squares report the original human-subject results, while red points report the LLM replication. As in Figure~\ref{fig:F9}, each panel corresponds to one experimental module, and diversity is operationalized using four metrics derived from group members’ demographics, dispositions, experience, and political attitudes. Point estimates are cell means with 90\% bootstrapped confidence intervals. Here dissensus is measured across members' reports within a group.

For human groups, diversity has a modest association with disagreement. Prospect-theory groups show slightly higher dissensus in more diverse hierarchical settings, while the other modules change little. LLM point estimates fluctuate more across diversity levels and carry substantially wider confidence intervals. The estimates therefore indicate greater instability in measured dissensus among LLM groups, with limited precision about its direction.

Taken together, the diversity analysis shows little systematic relationship between group composition and average bias. Demographic, attitudinal, and experiential heterogeneity produces no consistent attenuation or amplification across the three tasks. Group composition alone therefore supplies limited leverage for changing the average bias patterns in these simulations.

The human-silicon contrast lies in precision and stability. Human-group estimates change modestly across diversity levels. LLM-group estimates fluctuate more and carry wider intervals, especially for dissensus. Average bias remains directionally stable, while internal disagreement is measured with greater uncertainty. This distinction reconciles the stable average effects in Figure~\ref{fig:F9} with the variable dissensus estimates in Figure~\ref{fig:F10}.

\subsubsection{Disagreement in Reported Final Decisions}\label{app:matched_dissensus}

Following \citet[p.~535, n.~74]{Kertzer_2022_Hawkish}, we measure dissensus as the sample variance of members' final responses within each group. A unanimous set of reports has variance zero. In the original questionnaire (supplement, pp.~A1--A4 and A7--A8), human horizontal-group members first indicated whether they thought their group had reached unanimity. Those answering yes reported the group's decision, while those answering no reported their own preference. In hierarchical groups, the leader reported a decision and advisers reported their own judgments. Silicon members completed individual final surveys. The horizontal comparison therefore includes both agreement about what the group decided and agreement in personal judgments. These end-of-discussion measures do not identify which members changed their minds.

For this exploratory comparison, we paired human and silicon groups with the same participant roster, task, decision structure, group size, and treatment. Roster matching uses recorded participant identifiers; persona profiles were sampled separately. The available simulation runs cover 757 groups across three tasks. Of the 1,619 group--task records eligible after matching and requiring complete human outcomes, 879 had a codable final answer for every silicon member. We accepted a tagged policy choice or a response on the benchmark's seven-category scale and excluded incomplete response sets, incompatible scales, and ambiguous answers. We calculated variances on a zero-to-one scale and gave each retained group equal weight within its task and structure. Coverage was lowest for intentionality bias, where 133 of 495 eligible pairs were retained. Estimates describe groups meeting these criteria.

\begin{table}[htbp]
\centering
\caption{Post-deliberation disagreement in matched human and silicon groups}
\label{tab:matched_dissensus}
\small
\setlength{\tabcolsep}{3.5pt}
\begin{tabular}{llrrrrc}
\toprule
Task & Structure & Pairs$^{a}$ & Human & Silicon & Gap & 95\% CI for gap \\
\midrule
Prospect theory & Horizontal & 224/371 & 0.130 & 0.203 & $0.074$ & $[0.049, 0.098]$ \\
Prospect theory & Hierarchical & 211/343 & 0.142 & 0.178 & $0.036$ & $[0.011, 0.062]$ \\
Intentionality bias & Horizontal & 56/221 & 0.037 & 0.021 & $-0.017$ & $[-0.033, -0.003]$ \\
Intentionality bias & Hierarchical & 77/274 & 0.036 & 0.015 & $-0.021$ & $[-0.030, -0.012]$ \\
Reactive devaluation & Horizontal & 138/181 & 0.042 & 0.013 & $-0.029$ & $[-0.040, -0.018]$ \\
Reactive devaluation & Hierarchical & 173/229 & 0.042 & 0.012 & $-0.029$ & $[-0.038, -0.022]$ \\
\bottomrule
\end{tabular}
\par\smallskip
\begin{minipage}{\textwidth}\small
$^{a}$ Retained pairs / eligible pairs. Human and silicon entries are mean within-group variances. The gap is silicon minus human. Intervals are pointwise percentiles from 10,000 bootstrap samples of matched pairs drawn within treatment and group-size strata. Each interval describes variation across groups for the recorded simulation runs.
\end{minipage}
\end{table}

\subsubsection{Qualitative Analyses of Agent Deliberation}
\label{app:deliberation}

We complement the quantitative comparisons with qualitative analyses of selected transcripts. We focus on three overarching patterns: anchoring on persona and preference persistence; information-seeking and cautious baselines; and the effect of group structure and aggregation rules on bias expression. For each, we present narrative summaries and representative excerpts.

\paragraph{Persona Anchoring and Preference Persistence in Prospect Theory}

Across simulations, agents frequently drew on their persona attributes when justifying their positions. Once they had articulated a stance, they rarely reversed it during discussion. Instead, deliberation tended to consolidate and strengthen initial leanings. This pattern is particularly visible in the prospect theory experiment under the gain frame, where agents choose between a safe option (Policy A, which guarantees that 200 lives will be saved) and a risky option (Policy B, which offers a one-third chance of saving all 600 lives and a two-thirds chance of saving none).

In one horizontal group with five advisors, four initially leaned toward the safe option and one toward the risky option. In their first-round messages, advisors made explicit reference to their risk attitudes and broader belief systems. One advisor described themselves as conservative and conscientious and therefore framed Policy A as the “more secure and responsible choice.” Another advisor, whose persona was described as comfortable with risk and supportive of militant internationalism, justified their support for Policy B by appealing to the possibility of saving everyone and “redefining our impact.” Over subsequent rounds, the majority repeatedly invoked arguments about certainty, responsibility, and the ethics of avoiding catastrophic failure, while the risk-seeking advisor continued to propose ways to mitigate the risks of Policy B, including international cooperation and resource mobilization.

Across multiple rounds, each advisor maintained their initial position. The four risk-averse advisors moved from tentative endorsements of Policy A to increasingly impatient demands to “finalize” and “move forward” with the safe choice. The risk-preferring advisor continued to advocate for Policy B and eventually accepted the group's choice of Policy A in the absence of a concrete strategy for lowering the probability of failure. Their final justification remained consistent with the persona and emphasized the lost opportunity for a more ambitious policy. The transcript illustrates how persona anchoring and preference persistence magnify initial differences and strengthen the starting positions.

\paragraph{Information-Seeking and Cautious Baselines in Intentionality Bias}

In the intentionality-bias experiment, the original study asked whether the presence of fatalities in a crisis scenario increases attributions of hostile intent. Human subjects tended to treat deaths as a powerful cue that an adversary acted purposefully. In our synthetic groups, agents reacted very differently. They repeatedly highlighted the phrase “without further information” in the treatment text and took it as a directive to avoid premature inference. The transcripts show that LLM agents interpret this linguistic cue as more salient than the emotional impact of fatalities.

In a representative horizontal group in the all-fatalities condition, every advisor began the discussion by emphasizing uncertainty. They raised mechanical failure and accidents as plausible causes and treated limited information as grounds for caution. The discussion quickly focused on process and intelligence gathering. In subsequent rounds, the advisors prioritized coordination with allies and specific channels for collecting reliable information while avoiding escalation. By the final round, all agents endorsed a cautious approach: they treated intentional hostility as “somewhat unlikely” or “unlikely,” and one advisor declined to choose a probability, citing insufficient evidence.

The final justifications show how LLM agents allocate attention. Advisors repeatedly cite the lack of concrete evidence and the escalation risk associated with speculation. They acknowledge the fatalities as tragic and assign them little diagnostic weight. Institutional mechanisms for intelligence sharing instead organize the discussion. This pattern helps explain the flat attribution response among synthetic subjects: informational caveats overshadow the emotional salience of fatalities.

\paragraph{Group Structure, Aggregation Rules, and Bias Expression}

The third qualitative pattern concerns how institutional design shapes group outcomes when individual attitudes are stubborn. Across both horizontal and hierarchical conditions, agents show relatively little within-subject attitude change during deliberation. Advisors and leaders alike rarely revise their core positions, even after acknowledging other participants’ arguments. Under these circumstances, group structure and aggregation rules become decisive.

In hierarchical groups, leaders seldom adjusted their evaluations in response to advisor input. Hawkish or skeptical advisors shifted the discussion toward tougher safeguards, while the leader's final judgment remained close to the initial position. In a reactive-devaluation scenario with a China-authored proposal, the leader framed the proposal as a pragmatic opportunity to reduce tensions, whereas several advisors stressed China's record on intellectual property and currency manipulation. Over many rounds, the group developed a more stringent enforcement framework with phased milestones, US-led oversight, and automatic penalties. The leader ultimately expressed stronger support than any advisor, and the group outcome mirrored the leader's prior position.

Horizontal groups lack a formal leader, so aggregation rules determine how individual answers map onto collective outcomes. Deliberation seldom alters initial preferences in a substantive way. In some prospect-theory groups, the distribution of final preferences remains unchanged from the first round, while majority choice, the median, and unanimity requirements produce different group-level outcomes. Deliberation in these simulations primarily justifies existing positions and coordinates decision rules.

Taken together, the transcripts indicate that group deliberation affects outcomes primarily through institutional channels. Leader authority shapes hierarchical groups, while aggregation rules shape horizontal groups. Stable personas limit within-agent attitude change.

\subsection{Selected Deliberation Transcript Summaries}
\label{app:transcripts}

This section summarizes three deliberations corresponding to the qualitative analyses above. Each summary follows the sequence of discussion rounds and preserves selected phrases from the agents' responses. ``Leader'' and ``Advisor'' labels indicate roles in hierarchical settings. The archived simulation outputs contain the complete transcripts.

\subsubsection{Prospect Theory, Gain Frame (Horizontal Group)}

\textbf{Condition: Prospect Theory – Gain Frame}

\textbf{Round 1.} Advisor 4 begins by stressing discomfort with risk and an inclination toward Policy~A because it guarantees that 200 lives will be saved. Advisor~1 also favors Policy~A for its certainty, arguing that “it is always better to ensure some lives are saved rather than risk losing all.” Advisor~2 emphasizes reliability and the importance of certainty in critical situations. Advisor~5 invokes their “conservative and conscientious nature” as a reason to support Policy~A. In contrast, Advisor~3 highlights their comfort with risk and “militant internationalism,” arguing that the chance to save all 600 lives may justify the risk and explicitly asking others for their views.

\textbf{Round 2.} Advisor~3 reopens the case for Policy~B, asking whether the potential gains might outweigh the risks. The other advisors acknowledge this perspective and reiterate their discomfort with a two-thirds chance of saving no one. Several propose exploring ways to mitigate the risks of Policy~B while maintaining that the guaranteed outcome of Policy~A remains more attractive “in such a critical situation.”

\textbf{Rounds 3–7.} Over successive rounds, the majority continues to argue for Policy~A, increasingly framing the choice in terms of responsibility and urgency. Advisors express a desire to ``finalize" the decision to ensure that 200 lives are saved. Advisor~3 repeatedly suggests collaborating with allies and humanitarian organizations to increase the success probability of Policy~B and calls for “innovative strategies” that could “redefine our impact.” The group ends without a concrete risk-reduction mechanism.

\textbf{Final decisions.} In the final justifications, Advisors~1, 2, 4, and 5 all choose Policy~A. They describe this as a “responsible” option that prioritizes certainty and minimizes the risk of catastrophic loss, often citing their conservative or risk-averse traits. Advisor~3 remains consistent with their persona and selects Policy~B, acknowledges the group's preference for certainty, and argues that “taking calculated risks with a collaborative approach could redefine our impact and maximize the potential benefits.” This transcript underpins the analysis of persona anchoring and preference persistence in Appendix~\ref{app:deliberation}.

\subsubsection{Intentionality Bias, All Fatalities (Horizontal Group)}

\textbf{Condition: Intentionality Bias – All Fatalities}

\textbf{Round 1.} All five advisors immediately focus on uncertainty about the cause of the sinking. They mention mechanical failure and accidental causes as plausible alternatives and caution against “jumping to conclusions” about intentional action. The fatalities receive acknowledgment and little diagnostic weight. Advisors stress the importance of collecting more information about the adversary's motives.

\textbf{Rounds 2–4.} The group quickly converges on a process-oriented strategy. Advisors propose prioritizing intelligence gathering, coordinating with international allies, and designing mechanisms to ensure that any response is based on “solid evidence.” They repeatedly emphasize the need to avoid unnecessary escalation. Phrases such as “we should remain cautious before jumping to conclusions” and “let’s stay open to new evidence” recur across agents.

\textbf{Final decisions.} In their final explanations, advisors mostly choose options indicating that hostile intent is “somewhat unlikely” or “unlikely,” with one advisor effectively abstaining by refusing to assign a probability in the absence of further information. Their written justifications stress the lack of concrete evidence, the complexity of the regional context, and the dangers of escalation. They treat intelligence gathering and collaboration with allies as the substantive policy response, relegating attribution of intent to a secondary question to be answered later. This transcript supports the analysis of information-seeking and cautious baselines in Appendix~\ref{app:deliberation}.

\subsubsection{Reactive Devaluation, China Authorship (Hierarchical Group)}

\textbf{Condition: Reactive Devaluation – China Authorship}

\textbf{Round 1.} The leader opens by inviting initial reactions to a Chinese-drafted proposal that includes mutual tariff reductions, intellectual property protections, currency adjustments, and the creation of a UN watchdog agency. Advisors immediately express skepticism, focusing on China’s historical record on transparency, intellectual property enforcement, and currency manipulation. The overarching theme is distrust, with advisors asking what guarantees exist that China will follow through.

\textbf{Rounds 2–5.} Guided by the leader, the group gravitates toward the enforcement problem. Discussion centers on whether the proposed UN watchdog could provide credible oversight. The leader tends to see the watchdog as a promising solution, whereas several advisors view it as weak and potentially exploitable. One advisor argues that UN bodies “struggle to enforce rules against powerful nations”; another questions their impartiality. Through deliberation, the group gradually develops the idea of a phased agreement, in which concessions such as tariff reductions are conditional on meeting verified milestones in intellectual property protection and currency policy.

\textbf{Rounds 6–11.} The group refines this phased framework, specifying early milestones focused on intellectual property and currency, suggesting independent audits, and insisting on US-led oversight. Advisors push for automatic penalties, such as immediate tariff reinstatement, for non-compliance. The leader repeatedly summarizes and endorses this framework, characterizing it as a “balanced” and “pragmatic” way to move forward. Despite this convergence on institutional safeguards, underlying attitudes toward China remain polarized: some advisors continue to describe China as fundamentally untrustworthy and argue for extreme caution.

\textbf{Final decisions.} In their final justifications, advisors assign moderate support scores, combining recognition of the proposal's potential with lingering doubt about enforceability and China's reliability. The leader, by contrast, expresses relatively strong support, emphasizing that the phased framework with US-led oversight and automatic penalties “mitigates risks effectively” and offers a “cautious yet actionable path forward.” Deliberation transforms the imagined regulatory architecture and leaves deep skepticism among advisors, while the group outcome mirrors the leader's more optimistic prior position. Institutional design and leader authority determine the final group decision, as discussed in Appendix~\ref{app:deliberation}.

\clearpage
\subsection{Shrinkage PPI and Estimated-$\lambda$ Diagnostics}
\label{app:shrinkage_ppi}

The full PPI correction sets $\lambda=1$, using the entire LLM-implied adjustment. A fixed shrinkage extension uses a tuning parameter $\lambda\in[0,1]$; we also evaluate a weight estimated from the calibration sample:
\[
\theta_{\lambda}=\theta(Y_{\mathrm{cal}})+\lambda\bigl[\theta(\hat{Y}_{\mathrm{all}})-\theta(\hat{Y}_{\mathrm{cal}})\bigr].
\]
At $\lambda=0$ the estimator reduces to the human calibration sample; at $\lambda=1$ it recovers vanilla PPI. An optimal control-variate weight can be estimated from the calibration sample as
\[
\hat\lambda = \frac{\mathrm{Cov}(\phi_Y,\,\phi_{\hat Y})}{\mathrm{Var}(\phi_{\hat Y})},
\]
where $\phi_Y$ and $\phi_{\hat Y}$ are the unit-level ATE influence functions for the human and LLM responses, respectively \citep{angelopoulos2023ppi}.

Table~\ref{tab:lambda_ranking} reports average absolute error across modules and calibration shares for the narrative+direct configuration. Vanilla PPI ($\lambda=1$) has the highest error among the calibrated methods, while feasible methods favor small $\lambda$ values. Median estimated $\hat\lambda$ values range from approximately 0.11 to 0.12 across calibration shares. At 10\% calibration, estimated-$\lambda$ PPI and human-only calibration have similar mean error of approximately 0.150. Estimated-$\lambda$ PPI improves on human-only calibration by less than 0.001 in mean absolute error at each reported share. Fixed $\lambda=0.25$ improves error by about 0.001 at 25\% calibration and is essentially tied with human-only estimation at 50\% and 75\%. These results support limited shrinkage toward the silicon prediction.

\begin{table}[htbp]
\centering
\small
\caption{Shrinkage PPI performance by method and calibration share (narrative+direct)}
\label{tab:lambda_ranking}
\begin{tabular}{llcc}
\toprule
Calibration \% & Method & Mean error & Median $\lambda$ \\
\midrule
\multirow{5}{*}{10\%}
 & Estimated $\hat\lambda$ & 0.150 & 0.11 \\
 & Human-only calibration & 0.151 & 0.00 \\
 & $\lambda=0.25$ & 0.151 & 0.25 \\
 & $\lambda=0.50$ & 0.158 & 0.50 \\
 & Vanilla PPI ($\lambda=1$) & 0.184 & 1.00 \\
\midrule
\multirow{5}{*}{25\%}
 & $\lambda=0.25$ & 0.085 & 0.25 \\
 & Estimated $\hat\lambda$ & 0.085 & 0.12 \\
 & Human-only calibration & 0.086 & 0.00 \\
 & $\lambda=0.50$ & 0.087 & 0.50 \\
 & Vanilla PPI ($\lambda=1$) & 0.100 & 1.00 \\
\midrule
\multirow{5}{*}{50\%}
 & Estimated $\hat\lambda$ & 0.048 & 0.12 \\
 & $\lambda=0.25$ & 0.049 & 0.25 \\
 & Human-only calibration & 0.049 & 0.00 \\
 & $\lambda=0.50$ & 0.051 & 0.50 \\
 & Vanilla PPI ($\lambda=1$) & 0.060 & 1.00 \\
\midrule
\multirow{5}{*}{75\%}
 & Estimated $\hat\lambda$ & 0.028 & 0.12 \\
 & Human-only calibration & 0.028 & 0.00 \\
 & $\lambda=0.25$ & 0.028 & 0.25 \\
 & $\lambda=0.50$ & 0.030 & 0.50 \\
 & Vanilla PPI ($\lambda=1$) & 0.035 & 1.00 \\
\bottomrule
\end{tabular}

\begin{minipage}{0.95\linewidth}
\vspace{0.5em}
\small Note: Mean absolute error across the three experimental modules (intentionality bias, prospect theory, reactive devaluation) for the narrative+direct LLM configuration. Estimated $\hat\lambda$ uses the unconstrained control-variate weight from the calibration sample; a variant clipped to $[0,1]$ is evaluated separately. Errors are averaged over resampling draws and tasks on their recorded outcome scales. Source: \texttt{ppi\_lambda\_comparison\_feasible\_ranking.csv}.
\end{minipage}
\end{table}

\subsection{Biased-Sample Calibration Test}
\label{app:biased_sample}

A separate test asks whether LLM-based PPI can correct compositionally biased human calibration samples. For human samples biased along observable covariates (e.g., overrepresenting conservatives, males, or highly educated respondents), the estimator combines the full-sample synthetic estimate with a residual correction from the biased sample. Its ability to recover the full human benchmark depends on how well that residual correction transfers across the sampling imbalance:
\[
\hat\theta_{\mathrm{PPI}}^{\mathrm{biased}} = \hat\theta(\hat{Y}_{\mathrm{all}}) + \bigl[\hat\theta(Y_{\mathrm{biased}}) - \hat\theta(\hat{Y}_{\mathrm{biased}})\bigr].
\]

We test this using 16 bias scenarios (ideology, risk attitude, education, income, age, political interest, gender, and race, each skewed high or low) across three modules and calibration shares of 5\%, 10\%, 25\%, 50\%, and 75\%.

Table~\ref{tab:biased_sample} and Figure~\ref{fig:biased_sample_comparison} report the aggregate results. Human calibration consistently improves on raw LLM estimates as the calibration share increases. To assess the information in the silicon adjustment, we compare the sign of the full-minus-subsample silicon treatment effect with the sign of the corresponding human difference for each draw. The signs agree in 53.9\% of draws. Adding the silicon difference to the human subsample estimate brings it closer to the full human estimate in 40.7\% of draws. The table averages within-scenario median absolute errors across tasks and bias scenarios; PPI has higher error by this criterion at every calibration share. Means calculated across the underlying draws likewise favor human-only estimation in every task-by-calibration-share cell. The win rate and the error summaries answer different questions about the distribution of performance.

\begin{table}[htbp]
\centering
\small
\caption{Biased-sample PPI performance (narrative+direct, aggregated across modules and bias scenarios)}
\label{tab:biased_sample}
\begin{tabular}{rcccc}
\toprule
Cal.\ \% & Human err & PPI err & PPI$-$Human & PPI beats raw \% \\
\midrule
5  & 0.185 & 0.221 & $+$0.036 & 48.3 \\
10 & 0.129 & 0.155 & $+$0.026 & 62.0 \\
25 & 0.075 & 0.091 & $+$0.015 & 79.7 \\
50 & 0.045 & 0.054 & $+$0.009 & 89.9 \\
75 & 0.027 & 0.034 & $+$0.006 & 97.3 \\
\bottomrule
\end{tabular}

\begin{minipage}{0.95\linewidth}
\vspace{0.5em}
\small Note: ``Human err'' averages the within-scenario median absolute ATE errors of the biased human calibration sample relative to the full human benchmark. ``PPI err'' applies the same aggregation to PPI-calibrated errors. ``PPI beats raw'' reports the share of simulation draws where PPI error is smaller than raw LLM error. Each task-by-scenario cell contains 500 simulation draws. Error medians and draw-level win rates are calculated within cells, then averaged across three tasks and 16 bias scenarios. Source: \texttt{ppi\_biased\_sample\_aggregate\_summary.csv}.
\end{minipage}
\end{table}

One possible explanation for PPI underperformance is modest treatment-effect heterogeneity across the biasing covariates. Similar subgroup-specific ATEs would keep biased calibration samples close to the full human benchmark, leaving little composition bias for silicon predictions to address. This interpretation calls for direct examination of subgroup effects and residual transport across sampling scenarios. The aggregate heatmap appears in Figure~\ref{fig:biased_sample_comparison}.

\subsection{LLM Configuration Comparison}
\label{app:config_comparison}

Table~\ref{tab:config_ranking} ranks the four prompt-by-reasoning configurations by calibrated treatment-effect error, averaged across modules, calibration shares, and shrinkage methods. Narrative+direct produces the lowest calibrated error, followed by QA+CoT and narrative+CoT. Narrative+CoT performs best on raw LLM error, and this advantage disappears after calibration. The ranking supports the main-text choice of narrative+direct as the primary calibrated configuration.

\begin{table}[htbp]
\centering
\small
\caption{Configuration ranking by calibrated treatment-effect error}
\label{tab:config_ranking}
\begin{tabular}{llcc}
\toprule
Configuration & Best method & Mean error & Median error \\
\midrule
narrative+direct & est.\ $\hat\lambda$ & 0.078 & 0.067 \\
QA+CoT & est.\ $\hat\lambda$ (clip) & 0.097 & 0.085 \\
narrative+CoT & $\lambda=0.25$ & 0.098 & 0.083 \\
QA+direct & est.\ $\hat\lambda$ (clip) & 0.107 & 0.091 \\
\bottomrule
\end{tabular}

\begin{minipage}{0.95\linewidth}
\vspace{0.5em}
\small Note: Each row reports the best feasible calibration method for that configuration, averaged over three modules and four calibration shares (10\%, 25\%, 50\%, 75\%). Source: \texttt{overall\_config\_method\_ranking.csv}.
\end{minipage}
\end{table}

\end{document}